\documentclass[amsmath,amssymb,showkeys,superscriptaddress,floatfix,aps,prd,10pt]{revtex4}
\usepackage{graphicx,epstopdf}
\usepackage[caption=false]{subfig}
\usepackage{multirow}
\usepackage{appendix}
\usepackage{xcolor}
\usepackage[colorlinks=true, urlcolor=blue, linkcolor=blue, citecolor=green]{hyperref}

\def\xslash{x\!\!\!\slash }

\def\xslash{x\!\!\!\slash }

\def\beq{\begin{equation}}
\def\eeq{\end{equation}}
\def\bea{\begin{eqnarray}}
\def\eea{\end{eqnarray}}
\def\beeq{\begin{eqnarray}}
\def\eeeq{\end{eqnarray}}

\def\vel{\left|}
\def\ver{\right|}

\def\ba{\begin{array}}
\def\ea{\end{array}}

\def\xis0{{\Xi^{*0}}}

\def\g5{\gamma_5}

\begin{document}

\title{Exploring the magnetic dipole moments of $T_{QQ \bar q \bar s}$ and $T_{QQ \bar s \bar s}$  states  in the framework of QCD light-cone sum rules}
\author{K. Azizi}
\email[]{kazem.azizi@ut.ac.ir}
\affiliation{Department of Physics, University of Tehran, North Karegar Avenue, Tehran
14395-547, Iran}
\affiliation{Department of Physics, Do\v{g}u\c{s} University, Dudullu-\"{U}mraniye, 34775
Istanbul,  T\"{u}rkiye}
\author{U. \"{O}zdem}%
\email[]{ulasozdem@aydin.edu.tr}
\affiliation{ Health Services Vocational School of Higher Education, Istanbul Aydin University, Sefakoy-Kucukcekmece, 34295 Istanbul, T\"{u}rkiye}

\date{\today}
 
\begin{abstract}
 Motivated by the recent observation of the tetraquark $ T_{cc}^{+}$,  we investigate the magnetic dipole moments of the  possible single and double strange partners, $T_{QQ \bar q \bar s}$ and $T_{QQ \bar s \bar s}$,  with the spin-parity $ J^{P} = 1^{+}$  by means of the QCD light-cone sum rules. To this end, we model these states as diquark-antidiquark states with different organizations and  interpolating currents. The results of magnetic dipole moments obtained using different diquark-antidiquark structures differ from each other, considerably.
 The magnetic dipole moment is the leading-order response of a bound system to a soft external magnetic field. Therefore, it provides an excellent platform for investigation of  the inner structures of hadrons  governed by the quark-gluon dynamics of QCD.
\end{abstract}

\maketitle

\section{Introduction}\label{introduction}

With progress in the facilities and  accumulation of more experimental data, more and more multiquark (exotic) states are observed. In the case of tetraquarks, the most popular interpretations of the internal structures  are the  two-ordinary-meson molecular  as well as compact diquark-antidiquark configurations. However, it is still difficult to accurately specify  the inner structures of these states~(see for instance \cite{Esposito:2014rxa,Esposito:2016noz,Olsen:2017bmm,Lebed:2016hpi,Nielsen:2009uh,Brambilla:2019esw,Agaev:2020zad,Chen:2016qju,Ali:2017jda,Guo:2017jvc,Liu:2019zoy,Yang:2020atz,Dong:2021juy,Dong:2021bvy,Meng:2022ozq, Chen:2022asf} and references
therein, for further details).
In 2021, the LHCb Collaboration reported a narrow state in the $D^0 D^0 \pi^+$ invariant mass spectrum  just below the $D^0  D^{\ast +}$ mass threshold, namely $T^+_{cc}(3875)$ ($ T_{cc}^{+}$ for short)~\cite{LHCb:2021vvq,LHCb:2021auc}. Based on the analyses made, its valence quark content was suggested to be $cc\bar u \bar d$ and   its quantum numbers were estimated to be  $J^P = 1^+$. This resonance is the first observed doubly-charmed tetraquark state so far. This observation triggered many related theoretical and experimental studies and many  groups  performed analyses to figure out the spectroscopic parameters, decay modes, magnetic dipole moments, and production mechanisms of doubly-heavy tetraquark states including  $ T_{cc}^{+}$ via various approaches and models ~\cite{Du:2012wp,Qin:2020zlg,Feijoo:2021ppq,Deng:2021gnb,Yan:2021wdl,Wang:2021yld,Huang:2021urd,Agaev:2021vur,Chen:2021tnn,Jin:2021cxj,Ling:2021bir,Hu:2021gdg,Chen:2021cfl,Albaladejo:2021vln,Abreu:2021jwm,Du:2021zzh,Dai:2021vgf,Wang:2021ajy,Meng:2021jnw,Fleming:2021wmk,Xin:2021wcr,Ren:2021dsi,Albuquerque:2022weq, Azizi:2021aib,Ozdem:2021hmk,Kim:2022mpa,Abreu:2022lfy,Agaev:2022ast,Dai:2022ulk,Dai:2021wxi,Karliner:2017qjm,Eichten:2017ffp,Cheng:2020wxa,Braaten:2020nwp, Meng:2020knc,Dias:2011mi,Navarra:2007yw,Gao:2020ogo,Agaev:2020mqq,Agaev:2020zag,Agaev:2020dba,Agaev:2019lwh,Agaev:2018khe,Aliev:2021dgx,Mohanta:2020eed,Ke:2022vsi,Ozdem:2022yhi,Agaev:2022vhq, Wang:2022jop, Qin:2022nof, Wu:2022gie, Abreu:2022sra}.

The mass and  width of a particle are two main quantities that their values are provided by the experiments for the newly observed resonances. Most of theoretical studies calculate the mass and width (considering some dominant decays) of these states as well and compare the obtained results with the experimental data to get knowledge on the internal structure, quark content and quantum numbers of the particle. However, in the case of multiquark states, such comparisons lead to different quark-gluon structures for these states: In the case of tetraquarks, in most cases, both the molecular and compact tetraquark assignments lead to consistent results for the mass and width with the experimental data. This initiates many open questions about the internal organizations of these states. Therefore, more theoretical and experimental studies on the other quantities related to different interactions/decays of these states are needed to pin down the inner configurations of these resonances.  The electromagnetic properties of hadrons are among the main quantities that can provide extra information on the nature, structure and geometric shapes of hadrons.  The magnetic dipole moment (MDM) of hadrons is a  very important physical parameter in this regard. The MDMs of the hadrons are important measurable quantities, which carry substantial knowledge about their  underlying quark-gluon configurations, and can be used to
distinguish the suitable configurations  among  different theoretical assignments   and deepen our understanding of the underlying dynamics. Therefore,  it is important and  interesting to explore the MDMs of the newly discovered/suggested states.

In the present study, we evaluate the MDMs of the doubly-heavy single/double strange tetraquark partners of  $ T_{cc}^{+}$  in the diquark-antidiquark configuration using the QCD light-cone sum rule method \cite{Chernyak:1990ag, Braun:1988qv, Balitsky:1989ry}. We extend the calculations to include the corresponding bottom partners, $ T_{bb}$, as well. 
In this method an appropriate  correlation function is calculated in terms of hadronic parameters and their MDM on one side  and quark-gluon degrees of freedom and distribution amplitudes (DAs) of the on-shell photon on the other side. The DAs of photon are parameterized in terms of fuctions of different twists. Then,  Borel transform as well as continuum subtraction supplied by the quark-hadron duality assumption are applied to eliminate the  effects coming from the higher states and continuum. By matching the results obtained in two different sides, the sum rules for the MDM of the particles are obtained in terms of some auxiliary parameters (Borel mass squared $ M^2 $ and continuum threshold $ s_0 $), wave functions of the photon and various related parameters, quark masses, quark-gluon condensates etc.

This paper is structured in the following manner: In Sect. \ref{formalism}, QCD light-cone sum rules for the MDMs of $T_{QQ \bar q \bar s}$ and $T_{QQ \bar s \bar s}$ states (with  $Q=c, b$ and  $q=u,d$) are derived using  different interpolating currents. Numerical analysis of the MDMs is carried out in section \ref{numerical}, where their values are presented. The last section is reserved for a brief discussion and concluding remarks.


 \section{Formalism} \label{formalism}

 To obtain the MDMs of the particles under study, using the QCD light-cone sum rules, we consider the following two-point correlation function in the external background field:
\begin{equation}
 \label{edmn01}
\Pi _{\mu \nu }(p,q)=i\int d^{4}xe^{ip\cdot x}\langle 0|\mathcal{T}\{J_{\mu}(x)
J_{\nu }^{ \dagger }(0)\}|0\rangle_{\gamma}, 
\end{equation}%
where   $\gamma$ denotes the  background  electromagnetic field  and $J_{\mu}(x)$ stand for  the interpolating currents of the $T_{QQ}$ states with the spin-parity $ J^{P} = 1^{+}$. 
The interpolating current  $J_{\mu}(x)$ is one of the important quantities necessary to investigate the MDMs of the $T_{QQ}$ states using the QCD light-cone sum rule method. There are different ways to construct the  axial tetraquarks and corresponding interpolating currents using the  diquarks and  antidiquarks  of  different spin-parities.
For construction of the exotic states from the  diquarks and antidiquarks  in general, one may refer to the pioneering study by Jaffe \cite{Jaffe:2004ph}, where  the  light diquarks  are categorized as the scalar (good) and vector (bad) diquarks. In this study, the light vector  diquarks  are found  to be suppressed  in the spectroscopy of hadrons. Considering all the properties of the $T_{QQ \bar q \bar s}$ and $T_{QQ \bar s \bar s}$ states, the following diquark-antidiquark type  interpolating currents are constructed in Ref. \cite{Du:2012wp},  that we use to   calculate the MDMs of these states:
\begin{eqnarray}\label{curr1}
 &  J^1_{\mu}(x)= \big[Q^{aT}(x)C \gamma_\mu Q_b(x)][\bar q^a(x) \gamma_5 C \bar s^b(x)],\nonumber\\
 \nonumber\\
  &  J^2_{\mu}(x)= \big[Q^{aT}(x)C \gamma_5 Q_b(x)][\bar q^a(x) \gamma_\mu C \bar s^b(x)],\nonumber\\
  \nonumber\\
  &   J^3_{\mu}(x)= \big[Q^{aT}(x)C \sigma_{\mu\nu} \gamma_5 Q_b(x)][\bar q^a(x) \gamma^\nu C \bar s^b(x)],\nonumber\\
  \nonumber\\
  &   J^4_{\mu}(x)= \big[Q^{aT}(x)C \gamma^\nu \gamma_5 Q_b(x)][\bar q^a(x)  \sigma_{\mu\nu} C \bar s^b(x)],
     \end{eqnarray}
where   $q$ denotes the $u$, $d$ or $s$ quark in this equation. 
 The currents $J_\mu^1(x)$,  $J_\mu^2(x)$,   $J_\mu^3(x)$ and  $J_\mu^4(x)$ contain the heavy axial vector  diquak-light scalar antidiquark, heavy scalar diquak-light axial vector  antidiquark, heavy tensor  diquak-light axial vector  antidiquark and heavy vector diquak-light pseudotensor antidiquark, respectively.


The analysis of MDMs will begin with the calculation of the correlation function in terms of hadronic parameters. For this aim, we insert complete sets of hadronic states, which have the same quantum numbers with the considered interpolating currents, into the correlation function.
As a result, after performing the integral over four space-time,  we get
 \begin{align}
\label{edmn04}
\Pi_{\mu\nu}^{Had} (p,q) &= {\frac{\langle 0 \mid J_\mu (x) \mid
T_{QQ}(p, \varepsilon^\theta) \rangle}{p^2 - m_{T_{QQ}}^2}} 
\langle T_{QQ}(p, \varepsilon^\theta) \mid T_{QQ}(p+q, \varepsilon^\delta) \rangle_\gamma 
\frac{\langle T_{QQ}(p+q,\varepsilon^\delta) \mid {J^\dagger}_\nu (0) \mid 0 \rangle}{(p+q)^2 - m_{T_{QQ}}^2} \nonumber\\
&+\mbox{higher states}\,,
\end{align}
where $ p+q $ and $ p $ are the initial and final momenta of the tetraquark states and $\varepsilon^i  $ denote their polarization vectors.
To derive Eq. (\ref{edmn04}), we supposed that the hadronic representation of the sum rule is obtained by a single pole term in the initial and final states.  In the case of the multi-quark states such approximation should be verified by some supplementary arguments.  Since,  the  physical representation of the relevant sum rules receives contributions from the  two-hadron reducible terms as well (for details see for instance \cite{Weinberg:2013cfa,Lucha:2021mwx,Kondo:2004cr,Lee:2004xk,Lucha:2019pmp}). 
Hence, the two-meson contaminating effects have to be considered when extracting the parameters of the multi-quark exotic states. To this end, the quark propagator should be modified, i.e.,
\begin{align}
\frac{1}{m^{2}-p^{2}} \rightarrow \frac{1}{m^{2}-p^{2}-i\sqrt{p^{2}}\Gamma (p)%
},  \label{eq:Modif}
\end{align}%
where $\Gamma (p)$ is the finite width of the  exotic  states generated by the  two-meson contributions.  When these effects are properly considered in the sum rules,  they re-scale the residues of the  the  states under investigation leaving their  mass unchanged. Detailed analyses show that the two-meson contributions are small (these effects lead to additional  $\approx (5-7)\%$ uncertainty in the residue), and do not exceed uncertainties of the sum rules computations (see Refs. \cite{Wang:2015nwa,Agaev:2018vag,Sundu:2018nxt,Wang:2019hyc,Albuquerque:2021tqd,Albuquerque:2020hio,Wang:2020iqt,Wang:2019igl,Wang:2020cme}).  Hence,  it is expected that the results of MDMs are not affected by these effects  more and the contributions coming from these effects remain within the uncertainties of the results as well.  Therefore, one can safely neglect the contributions of the  two-meson states in the hadronic representation of the correlation function and  use the zero-width single-pole approximation.

The vacuum-hadron matrix elements of the final  and initial  currents, $\langle 0 \mid J_\mu(x) \mid T_{QQ}(p,\varepsilon^\theta) \rangle$ and $\langle T_{QQ}(p+q,\varepsilon^\delta) \mid {J^\dagger}_\nu (0) \mid 0 \rangle $ as well as the radiative transition matrix element, $\langle T_{QQ}(p,\varepsilon^\theta) \mid  T_{QQ} (p+q,\varepsilon^{\delta})\rangle_\gamma$ have the following forms in terms of the residues, polarization vectors,  invariant transition form factors and different Lorentz structures:
\begin{align}
\langle 0 \mid J_\mu(x) \mid T_{QQ}(p,\varepsilon^\theta) \rangle &= \lambda_{T_{QQ}} \varepsilon_\mu^\theta\,,\\
\langle T_{QQ}(p+q,\varepsilon^\delta) \mid {J^\dagger}_\nu (0) \mid 0 \rangle = &= \lambda_{T_{QQ}} \varepsilon_\nu^\delta\,,\\
\langle T_{QQ}(p,\varepsilon^\theta) \mid  T_{QQ} (p+q,\varepsilon^{\delta})\rangle_\gamma &= - \varepsilon^\tau (\varepsilon^{\theta})^\alpha (\varepsilon^{\delta})^\beta \bigg\{ G_1(Q^2) (2p+q)_\tau ~g_{\alpha\beta} 
+ G_2(Q^2) ( g_{\tau\beta}~ q_\alpha -  g_{\tau\alpha}~ q_\beta) 
\nonumber\\ &- \frac{1}{2 m_{T_{QQ}}^2} G_3(Q^2)~ (2p+q)_\tau 
q_\alpha q_\beta  \bigg\},\label{edmn06}
\end{align}
where $\varepsilon^\tau$ is the polarization of the photon,  $\lambda_{T_{QQ}}$ is residue of the doubly-heavy tetraquark state, and   $G_1(Q^2)$, $G_2(Q^2)$, and $G_3(Q^2)$ are the electromagnetic form factors with  $Q^2=-q^2$.

The next step is to combine  Eqs. (\ref{edmn04})-(\ref{edmn06}) to obtain the following form of the physical or hadronic side in terms of the related quantities:
\begin{align}
\label{edmn09}
 \Pi_{\mu\nu}^{Had}(p,q) &=  \frac{\varepsilon_\rho \, \lambda_{T_{QQ}}^2}{ [m_{T_{QQ}}^2 - (p+q)^2][m_{T_{QQ}}^2 - p^2]}
 \Big\{ G_2 (Q^2) 
 \Big(q_\mu g_{\rho\nu} - q_\nu g_{\rho\mu} -
\frac{p_\nu}{m_{T_{QQ}}^2}  \big(q_\mu p_\rho - \frac{1}{2}
Q^2 g_{\mu\rho}\big) 
  \nonumber\\
 &  
 +
\frac{(p+q)_\mu}{m_{T_{QQ}}^2}  \big(q_\nu (p+q)_\rho+ \frac{1}{2}
Q^2 g_{\nu\rho}\big) 
-  
\frac{(p+q)_\mu p_\nu p_\rho}{m_{T_{QQ}}^4} \, Q^2
\Big)
+\mbox{other independent structures}\Big\}.
\end{align}
The magnetic form factor, $F_M(Q^2)$, is related to  only  $G_2(Q^2)$ among the form factors,  
\begin{align}
\label{edmn07}
&F_M(Q^2) = G_2(Q^2)\,,
\end{align}
 and the MDM of the particles under study, $\mu_{T_{QQ}}$, is defined at static limit:
\begin{align}
\label{edmn08}
&\mu_{T_{QQ}} = \frac{ e}{2\, m_{T_{QQ}}} \,F_M(Q^2=0).
\end{align}
We will choose the structure $\varepsilon.p (p_\mu q_\nu- p_\nu q_\mu)$ to calculate the MDMs of the $T_{QQ}  $ states. This structure appears after a simple manipulation in the physical side presented above.

To get the QCD side of the correlation function, we insert the explicit forms of the interpolating currents presented in Eq. (\ref{curr1}) into Eq. (\ref{edmn01}), and contract the corresponding heavy and light quark fields with the help of Wick's theorem. The contraction results which are obtained in terms of the heavy and light quarks propagators for $T_{QQ \bar q \bar s}$ systems are different from those of the $T_{QQ \bar s \bar s}$ systems because of the number of identical quark fields. 
 For instance, the results of contractions in the case of $T_{QQ \bar q \bar s}$ particles  for the currents $J_\mu^1$ and  $J_\mu^3$ are obtained  as follows:
\begin{eqnarray}
\Pi _{\mu \nu }^{\mathrm{QCD}}(p,q)&=&i\int d^{4}xe^{ip\cdot x} \, \langle 0 \mid  
\Big\{\mathrm{%
Tr}\Big[ \gamma _{5} \widetilde S_{s}^{b^{\prime }b}(-x)\gamma _{5}S_{q}^{a^{\prime }a}(-x)\Big]    
\mathrm{Tr}\Big[ \gamma _{\mu }S_{Q}^{bb^{\prime
}}(x)  \gamma _{\nu } \widetilde S_{Q}^{aa^{\prime }}(-x)\Big]
\nonumber\\
&&
-\mathrm{%
Tr}\Big[ \gamma _{5} \widetilde S_{s}^{b^{\prime }b}(-x)\gamma _{5}S_{q}^{a^{\prime }a}(-x)\Big]    
\mathrm{Tr}\Big[ \gamma _{\mu }S_{Q}^{ba^{\prime
}}(x)  \gamma _{\nu } \widetilde S_{Q}^{ab^{\prime }}(-x)\Big]
\Big\} \mid 0 \rangle_{\gamma} ,  \label{QCDSide1}
\end{eqnarray}%
for the $J_\mu^1$ current, and 

\begin{eqnarray}
\Pi _{\mu \nu }^{\mathrm{QCD}}(p,q)&=&i\int d^{4}xe^{ip\cdot x} \, \langle 0 \mid 
\Big\{\mathrm{%
Tr}\Big[\sigma_{\mu\alpha} \gamma _{5} \widetilde S_{s}^{b^{\prime }b}(-x)\gamma _{5} \sigma_{\mu\alpha}S_{q}^{a^{\prime }a}(-x)\Big]    
\mathrm{Tr}\Big[ \gamma ^{\alpha }S_{Q}^{bb^{\prime
}}(x)  \gamma ^{\beta } \widetilde S_{Q}^{aa^{\prime }}(-x)\Big]
\nonumber\\
&&
-\mathrm{%
Tr}\Big[\sigma_{\mu\alpha} \gamma _{5} \widetilde S_{s}^{b^{\prime }b}(-x)\gamma _{5}\sigma_{\nu\alpha}S_{q}^{a^{\prime }a}(-x)\Big]    
\mathrm{Tr}\Big[ \gamma ^{\alpha }S_{Q}^{ba^{\prime
}}(x)  \gamma ^{\beta } \widetilde S_{Q}^{ab^{\prime }}(-x)\Big]
\Big\} \mid 0 \rangle_{\gamma} ,  \label{QCDSide2}
\end{eqnarray}%
for the $J_\mu^3$ current. Here, $S_{q}(x)$ and $S_{Q}(x)$ represent  the light and heavy-quark propagators, respectively.  They are given as
\begin{align}
\label{edmn12}
S_{q}(x)&=i \frac{{\xslash}}{2\pi ^{2}x^{4}} 
- \frac{\langle \bar qq \rangle }{12} \Big(1-i\frac{m_{q} \xslash}{4}   \Big)
- \frac{ \langle \bar qq \rangle }{192}m_0^2 x^2   
\Big(1-i\frac{m_{q} \xslash}{6}   \Big)
-\frac {i g_s }{32 \pi^2 x^2} ~G^{\mu \nu} (x) \Big[\rlap/{x} 
\sigma_{\mu \nu} +  \sigma_{\mu \nu} \rlap/{x}
 \Big],
\end{align}
and
\begin{align}
\label{edmn13}
S_{Q}(x)&=\frac{m_{Q}^{2}}{4 \pi^{2}} \Bigg[ \frac{K_{1}\Big(m_{Q}\sqrt{-x^{2}}\Big) }{\sqrt{-x^{2}}}
+i\frac{{\xslash}~K_{2}\Big( m_{Q}\sqrt{-x^{2}}\Big)}
{(\sqrt{-x^{2}})^{2}}\Bigg] 
-\frac{g_{s}m_{Q}}{16\pi ^{2}} \int_0^1 dv\, G^{\mu \nu }(vx)\Bigg[ \big(\sigma _{\mu \nu }{\xslash}
  +{\xslash}\sigma _{\mu \nu }\big)\nonumber\\
  &\times \frac{K_{1}\Big( m_{Q}\sqrt{-x^{2}}\Big) }{\sqrt{-x^{2}}}
+2\sigma_{\mu \nu }K_{0}\Big( m_{Q}\sqrt{-x^{2}}\Big)\Bigg],
\end{align}%
where $\langle \bar qq \rangle$ represents the light-quark  condensate, $m_0^2\langle \bar qq \rangle$ stands for the quark-gluon mixed condensate,  
$G^{\mu\nu}$ are the gluon field strength tensor, $v$ is the line variable,  and $K_0$, $K_1$ and  $K_2$ are the modified Bessel functions of the second kind.

 We do the calculations in $ x $ space  then move them to the momentum space by performing the related Fourier integrals. In terms of Feynman diagrams, our calculations are equivalent to calculations of some Feynman diagrams in the momentum space.  In Appendix A, we present these diagrams for some  lower-dimensional operators, although we take into consideration all the possible diagrams in the calculations.

The correlation functions obtained in Eqs. (\ref{QCDSide1}) and (\ref{QCDSide2}) contain different types of contributions:  Perturbative contributions, i.e., photon interacts with the light and heavy quark propagators perturbatively, and nonperturbative contributions, i.e., photon interacts with the light-quark fields at a large distance. 
In the case of the perturbative contributions, one of the light or heavy quark propagators in Eqs. (\ref{QCDSide1}) and (\ref{QCDSide2}) is replaced by
\begin{align}
\label{free}
S^{free}(x) \rightarrow \int d^4y\, S^{free} (x-y)\,\rlap/{\!A}(y)\, S^{free} (y)\,,
\end{align}
where $S^{free}(x)$ represents the first term of the light or heavy quark propagator, and the remaining propagators in Eqs. (\ref{QCDSide1}) and (\ref{QCDSide2}) are taken as full quark propagators.  
In the case of nonperturbative contributions, one of the light quark propagators in Eqs. (\ref{QCDSide1}) and (\ref{QCDSide2}), described the photon emission at large distances, is replaced through 
\begin{align}
\label{edmn14}
S_{\mu\nu}^{ab}(x) \rightarrow -\frac{1}{4} \big[\bar{q}^a(x) \Gamma_i q^b(x)\big]\big(\Gamma_i\big)_{\mu\nu},
\end{align}
 where  
 $\Gamma_i = I, \gamma_5, \gamma_\mu, i\gamma_5 \gamma_\mu, \sigma_{\mu\nu}/2$,
and the remaining propagators are taken as the full quark propagators.  When  Eq. (\ref{edmn14}) is used in computations of the nonperturbative contributions, we observe that matrix elements of the forms $\langle \gamma(q)\vel \bar{q}(x) \Gamma_i q(0) \ver 0\rangle$ and $\langle \gamma(q)\vel \bar{q}(x) \Gamma_i G_{\mu\nu}q(0) \ver 0\rangle$ are appeared. 
These matrix elements represent the DAs of the on-shell photon and are parameterized in terms of the wave function of the photon with different twists  (for details see Ref. \cite{Ball:2002ps}).  We should stress  that the photon can also be emitted at a long distance from the corresponding heavy quarks. However, due to the large mass of these  quarks, such contributions are suppressed. Therefore, these contributions are neglected in our computations. 
After the above-mentioned computations, the correlation function is obtained in terms of the wave functions of the photon and the parameters of the quarks and gluons and their interactions with the QCD vacuum. 

As a final step, we choose the structure $\varepsilon.p (p_\mu q_\nu- p_\nu q_\mu)$ from both representations of the correlation function and match its coefficients from both the hadronic and QCD sides. To eliminate the contributions of the higher states and continuum, we make use of the  Borel transformation as well as  continuum subtraction supplied by the quark-hadron duality assumption. As a result, we obtain the following sum rules: 
\begin{align}\label{sonj1}
 \mu^1_{T_{QQ}} &= \Delta_1 (M^2,s_0),~~~~~~~~ \mu^2_{T_{QQ}} = \Delta_2 (M^2,s_0),
 \end{align}
 \begin{align}\label{sonj2}
  \mu^3_{T_{QQ}} &= \Delta_3 (M^2,s_0),~~~~~~~~ \mu^4_{T_{QQ}} = \Delta_4 (M^2,s_0),
\end{align}
where $\mu^1_{T_{QQ}}$, $\mu^2_{T_{QQ}}$, $\mu^3_{T_{QQ}}$ and $\mu^4_{T_{QQ}}$ represent the sum rules for the MDMs obtained using the interpolating currents $J_\mu^1$, $J_\mu^2$, $J_\mu^3$ and $J_\mu^4$, respectively. For simplicity, in Appendix B, only the result for the $\Delta_1 (M^2,s_0)$ function is presented explicitly.


\section{Numerical results}\label{numerical}
In this section, we numerically analyze the MDMs of the $T_{QQ \bar q \bar s}$ and $T_{QQ \bar s \bar s}$ states. 
The expressions obtained for MDMs in the previous section contain various input parameters such as the masses of the quarks, quark, gluon and mixed condensates, photon's wave functions as well as the  masses and residues of the hadrons under study.    The input parameters are taken as: $m_u=m_d=0$, $m_s =93.4^{+8.6}_{-3.4}\,\mbox{MeV}$, $m_c = 1.27 \pm 0.02\,$GeV, 
$m_b = 4.18^{+0,03}_{-0.02}\,$GeV, 
$f_{3\gamma}=-0.0039$~GeV$^2$~\cite{Ball:2002ps},  
$\langle \bar ss\rangle $= $0.8 \langle \bar uu\rangle$ with 
$\langle \bar uu\rangle $=$(-0.24\pm0.01)^3\,$GeV$^3$~\cite{Ioffe:2005ym},  
$m_0^{2} = 0.8 \pm 0.1$~GeV$^2$~\cite{Ioffe:2005ym} and  $\langle g_s^2G^2\rangle = 0.88~ $GeV$^4$~\cite{Matheus:2006xi}. 
As mentioned, to obtain numerical values for the MDMs, we need the numerical values of the mass and residue of the $T_{QQ \bar q \bar s}$ and $T_{QQ \bar s \bar s}$ states as well. These values are borrowed from Ref.~\cite{Du:2012wp}. We also need the photon DAs and wave functions as well as the corresponding parameters: For completeness, we present their expressions in Appendix C.

As one can see from Eqs. (\ref{sonj1}) and (\ref{sonj2}) the sum rules depend also on the auxiliary  Borel and continuum threshold parameters $M^2$ and $s_0$. The choice of working windows for $M^2$ and $s_0$ has to fulfill the standard requirements of the method:  The pole contribution (PC) dominance and convergence of the operator product expansion (OPE). We require that the PC be larger than $\%30$ in the case of tetraquarks and the contribution of the higher dimensional nonperturbative operator (here dimension 8) be less than  $\%5$ of the total. To quantitatively define these constraints, it is suitable to employ the following formulas:
\begin{align}
 \mbox{PC} =\frac{\Delta (M^2,s_0)}{\Delta (M^2,\infty)},
 \end{align}
and 
\begin{align}
 \mbox{OPE Convergence} =\frac{\Delta^{Dim 8} (M^2,s_0)}{\Delta (M^2,s_0)},
 \end{align}
 where $\Delta^{Dim 8} (M^2,s_0)$ denotes the contribution of the highest dimensional term in the OPE. 
The working regions obtained for $M^2$ and $s_0$ as a result of these restrictions are given in Table \ref{parameter} together with PC and convergence of OPE values for each channel and current.  It follows from these values that the determined working regions for $M^2$ and $s_0$ meet the constraints coming from the dominance of PC and convergence of the OPE. 

\begin{widetext}
 
 \begin{figure}[htp]
\centering
\subfloat[]{\includegraphics[width=0.45\textwidth]{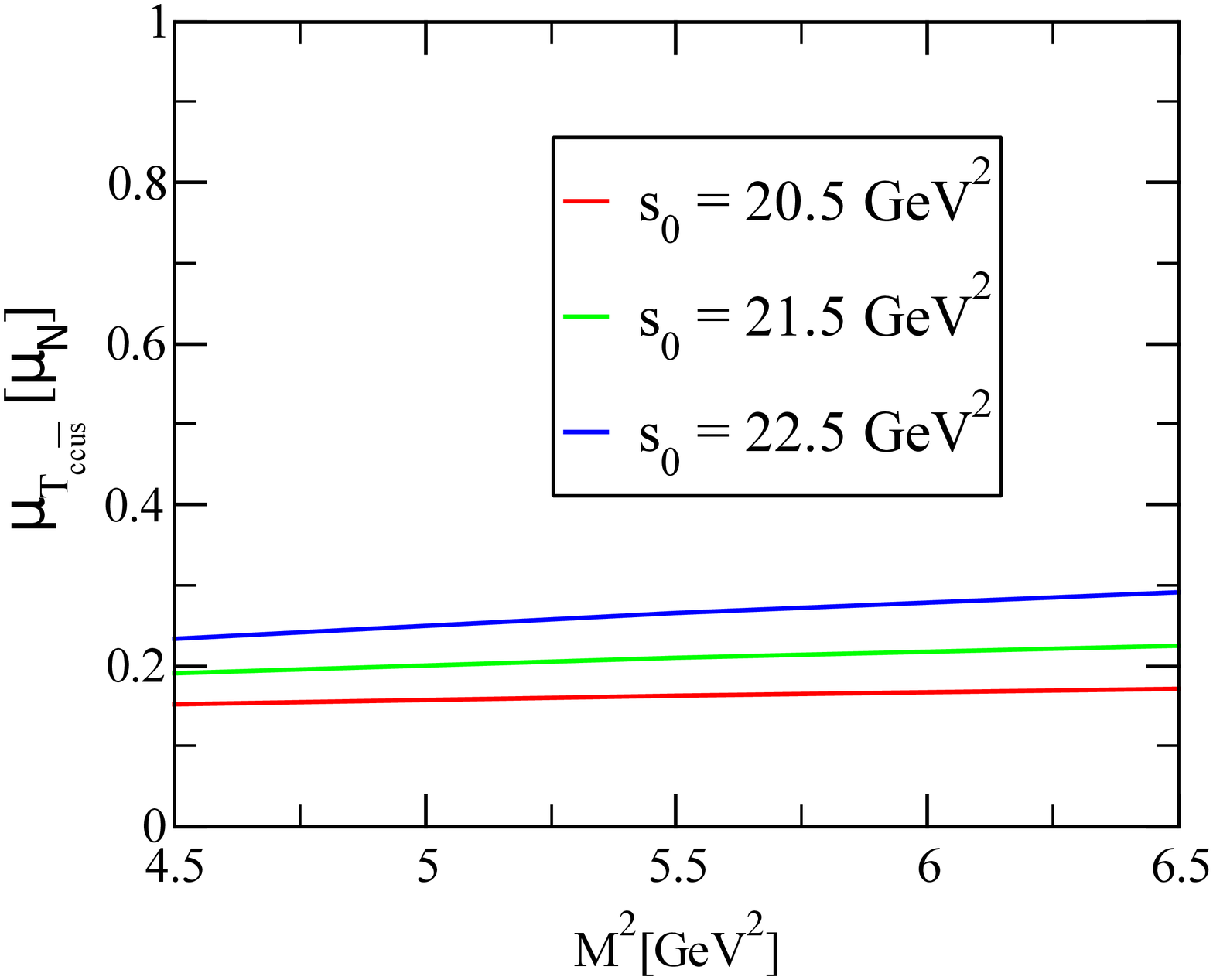}}
\subfloat[]{\includegraphics[width=0.45\textwidth]{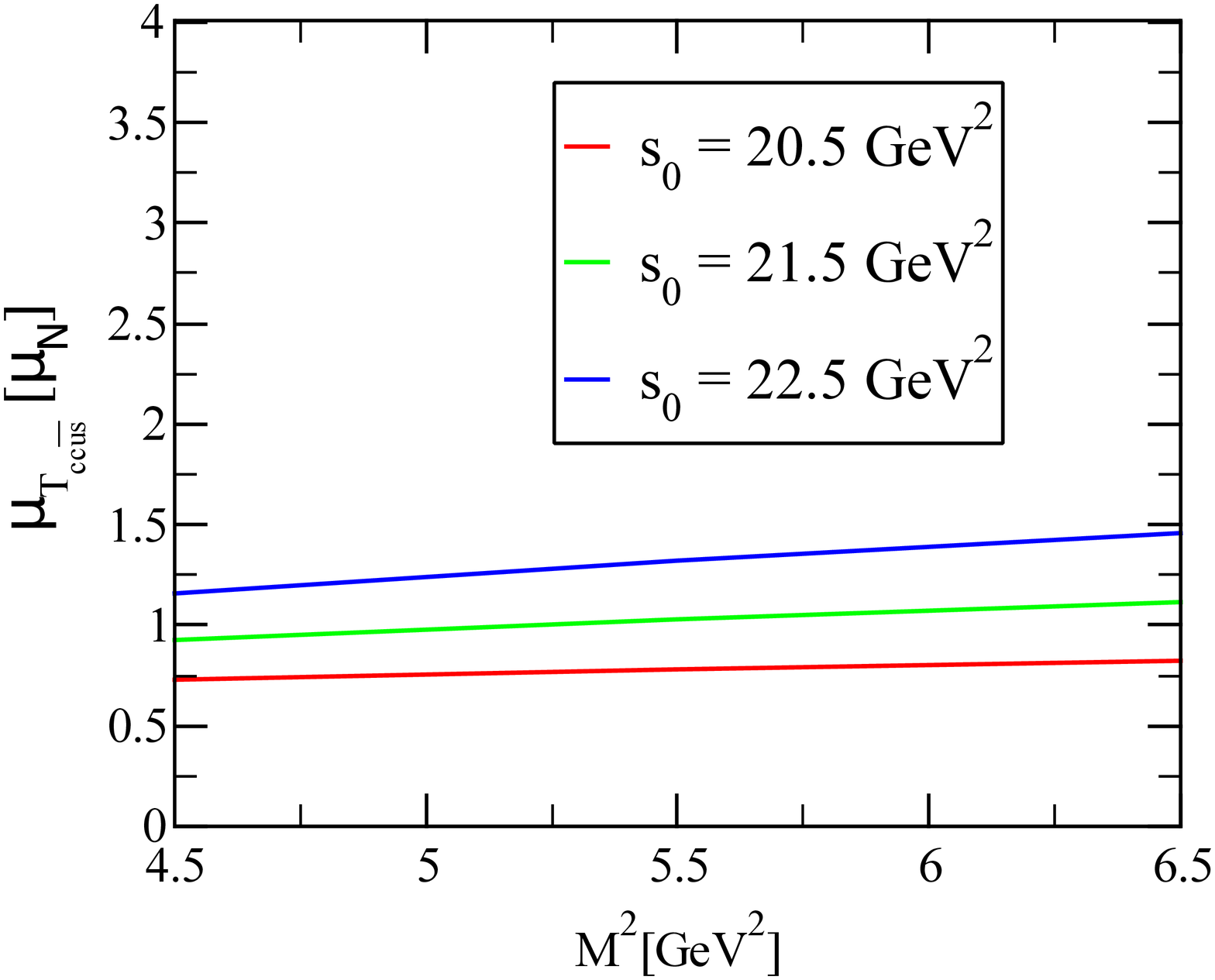}}\\
\subfloat[]{\includegraphics[width=0.45\textwidth]{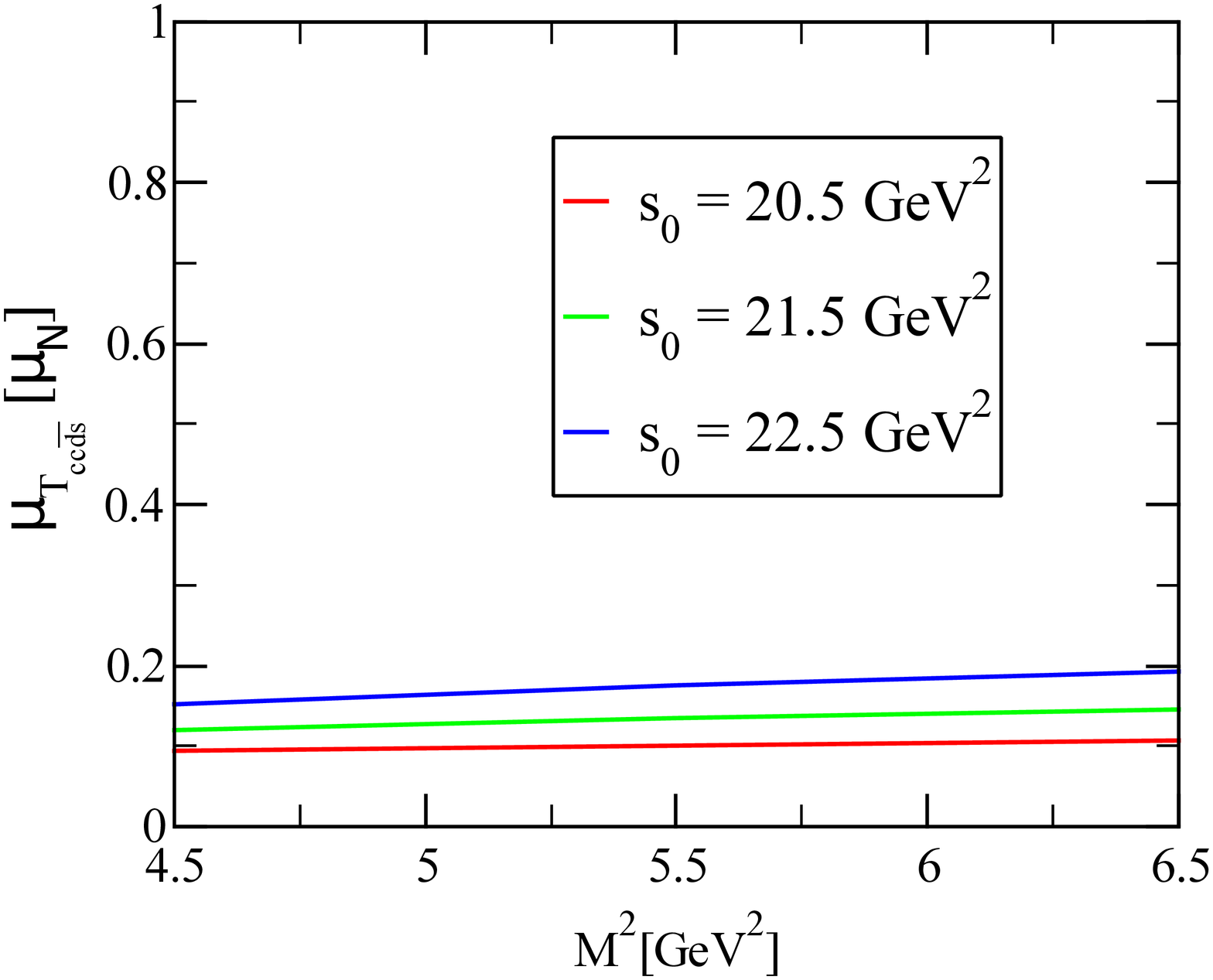}}
\subfloat[]{\includegraphics[width=0.45\textwidth]{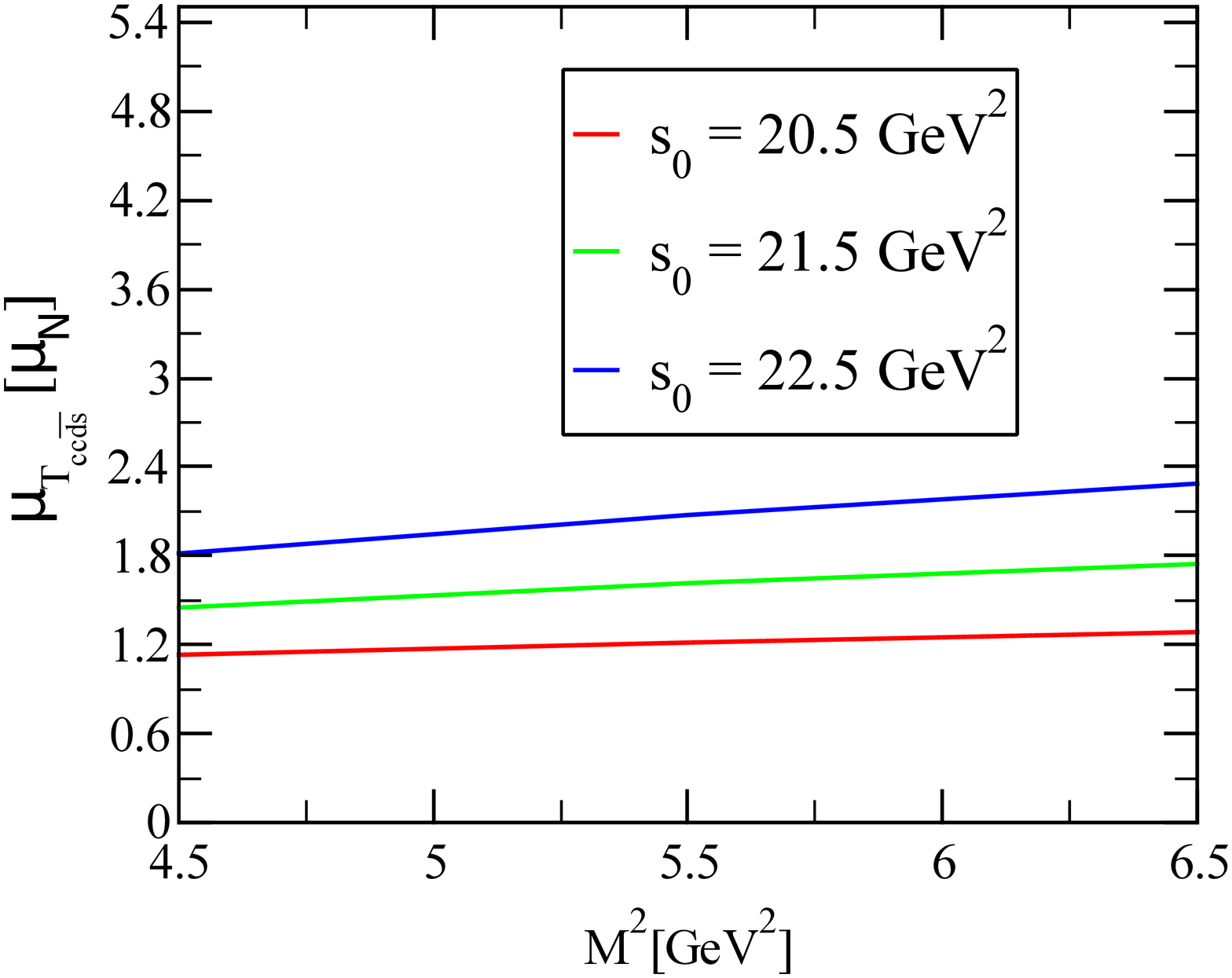}}\\
\subfloat[]{\includegraphics[width=0.45\textwidth]{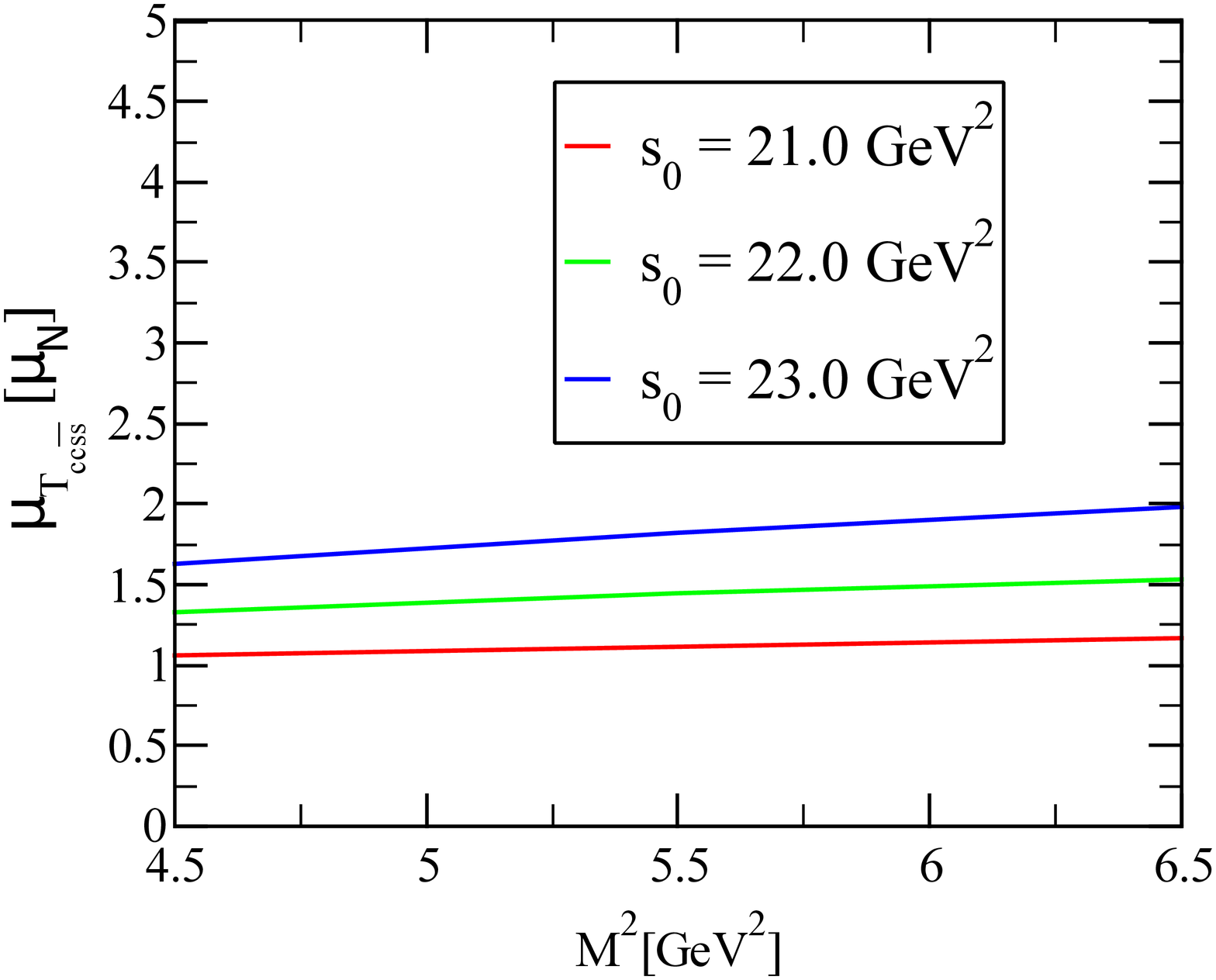}}
 \caption{Dependence of the MDMs of $T_{cc\bar q \bar s}$ and $T_{cc\bar s \bar s}$ tetraquark states on $M^2$ at three different values of $s_0$; (a) and (c) for the $J_{\mu}^1$ current; and (b), (d), and (e) for the $J_{\mu}^3$ current. }
 \label{Msqfig1}
  \end{figure}
  
  \end{widetext}

\begin{widetext}
 
 \begin{figure}[htp]
\centering
\subfloat[]{\includegraphics[width=0.45\textwidth]{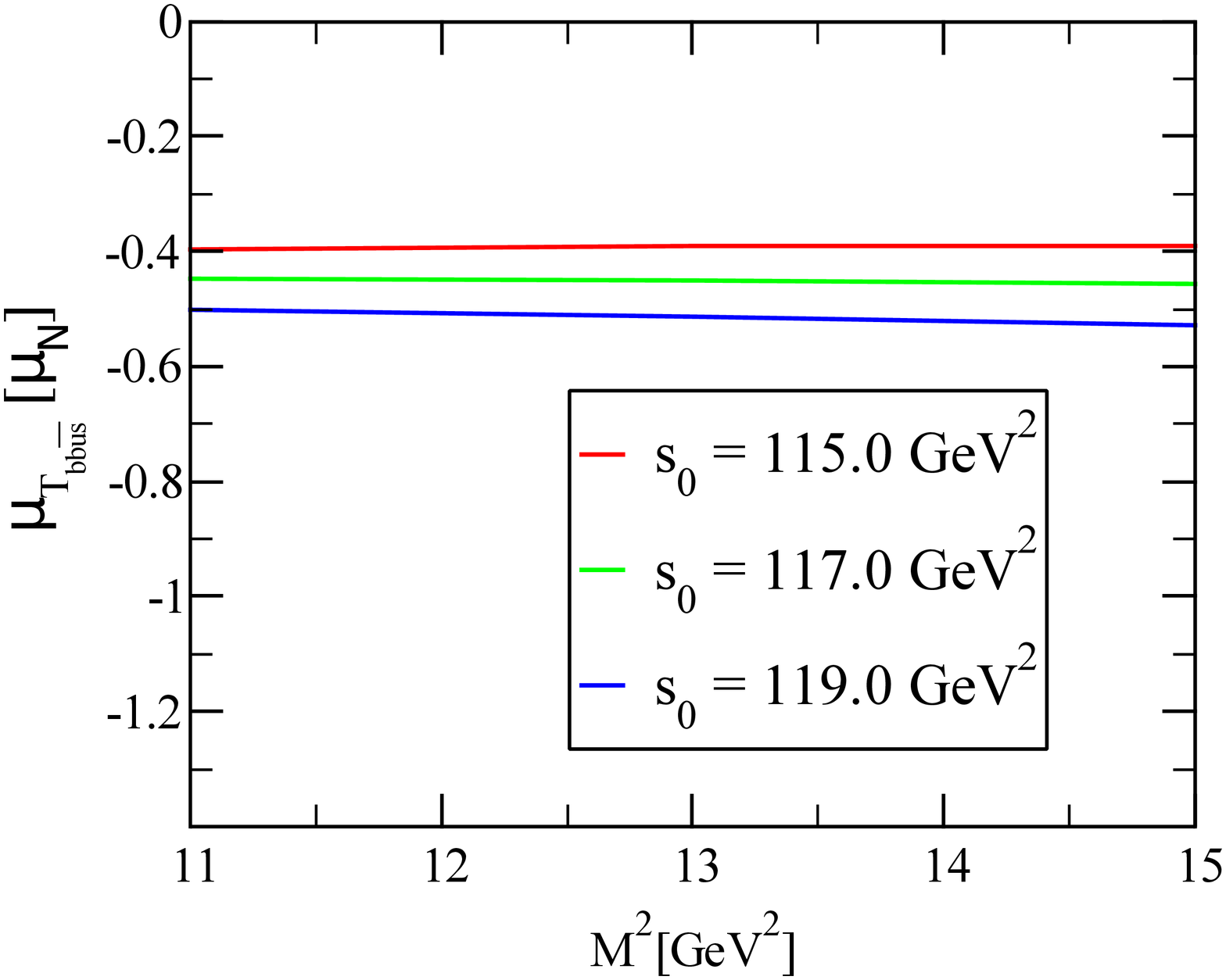}}
\subfloat[]{\includegraphics[width=0.45\textwidth]{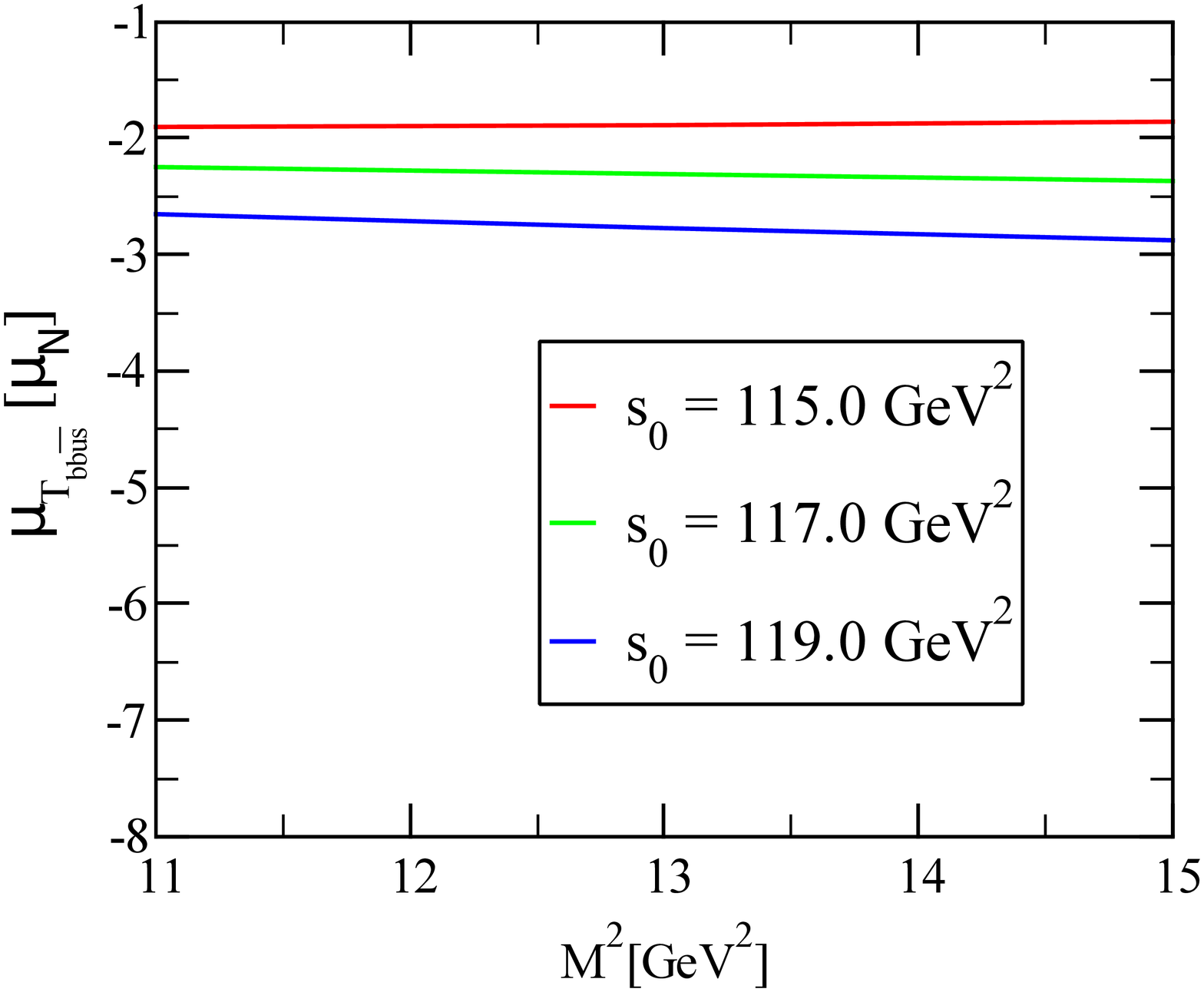}}\\
\subfloat[]{\includegraphics[width=0.45\textwidth]{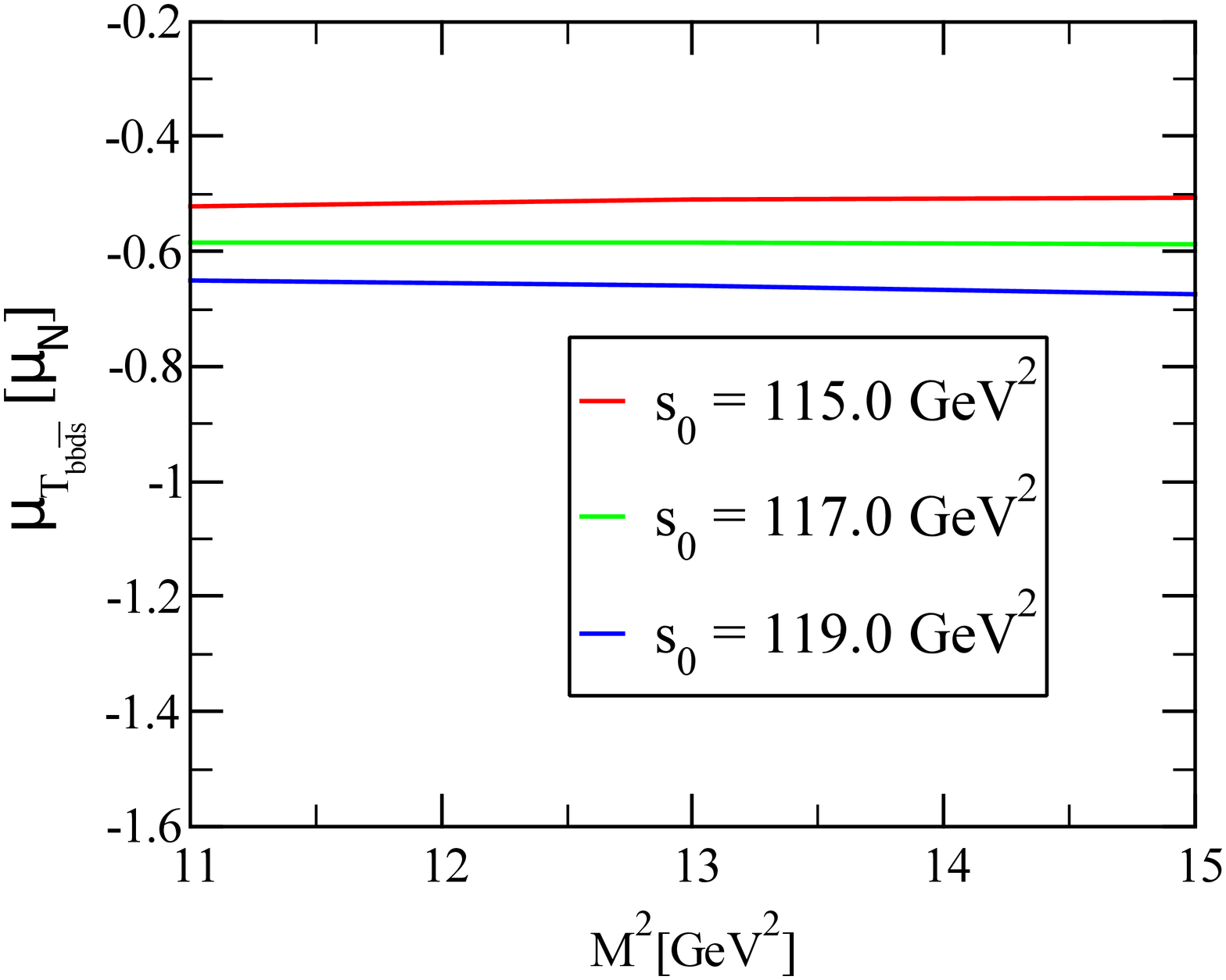}}
\subfloat[]{\includegraphics[width=0.45\textwidth]{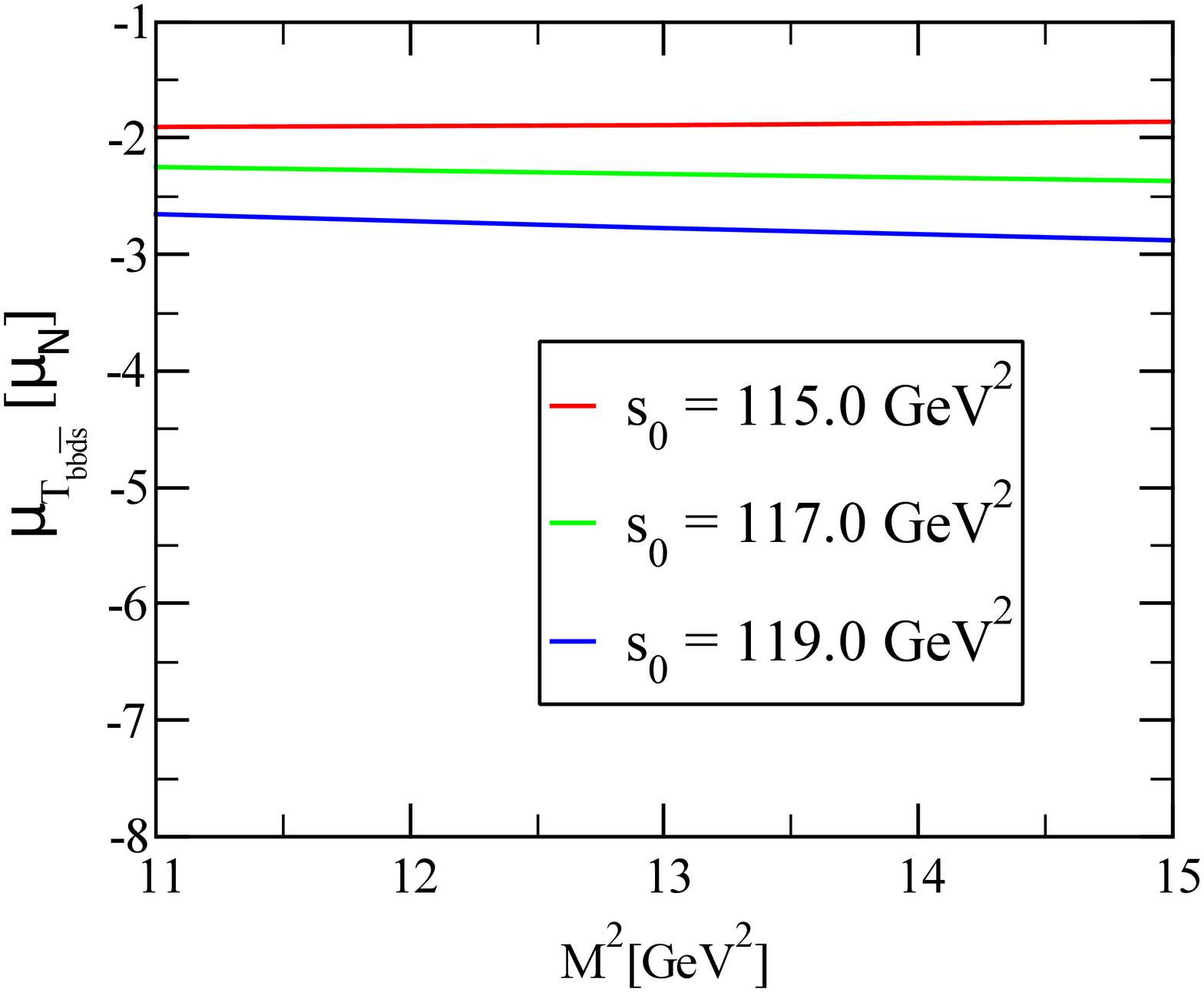}}\\
\subfloat[]{\includegraphics[width=0.45\textwidth]{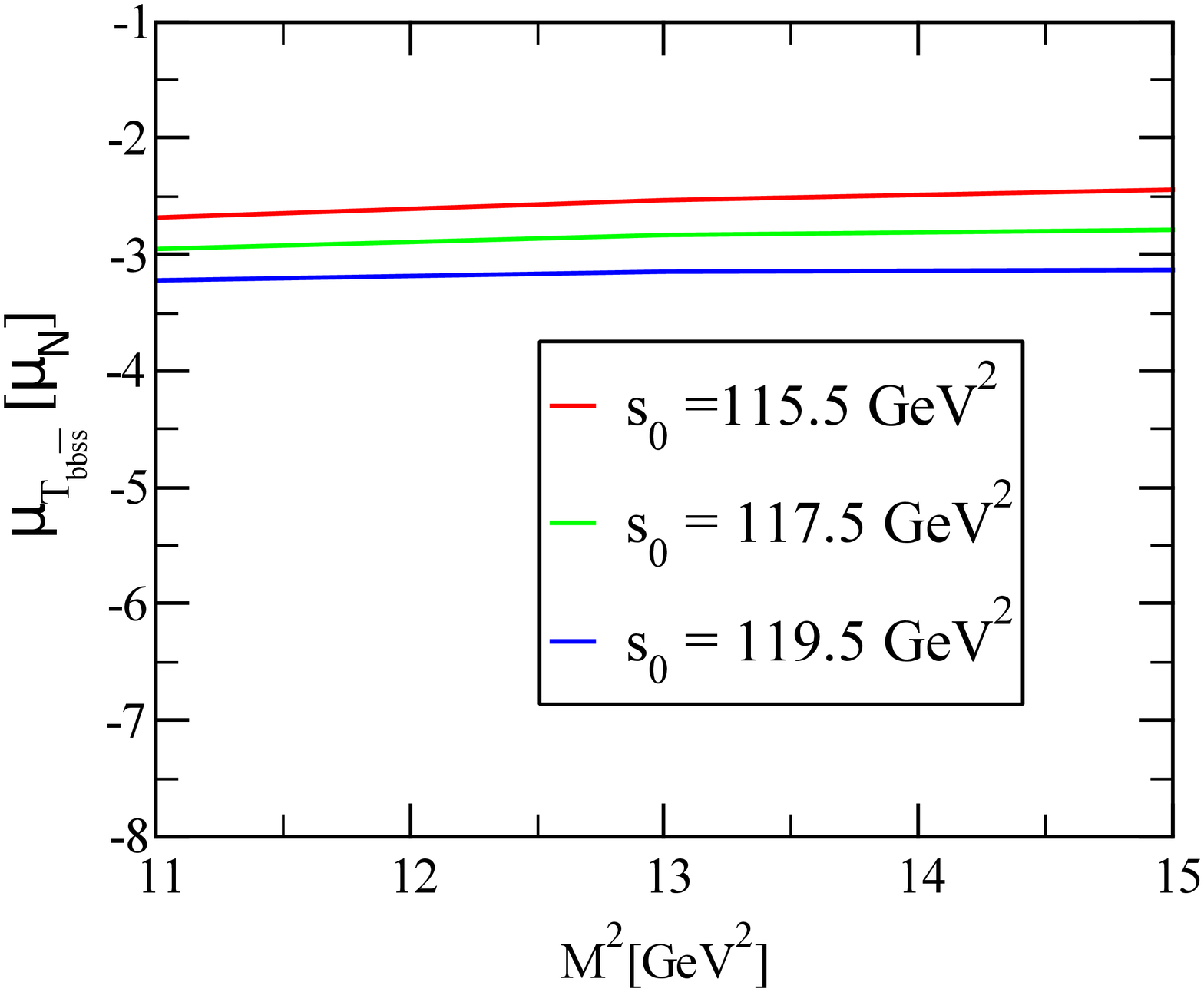}}
 \caption{Dependence of the MDMs of $T_{bb\bar q \bar s}$ and $T_{bb\bar s \bar s}$ tetraquark states on $M^2$ at three different values of $s_0$; (a) and (c) for the $J_{\mu}^1$ current; and (b), (d), and (e) for the $J_{\mu}^3$ current. }
 \label{Msqfig2}
  \end{figure}
  
  \end{widetext}

For completeness, for instance, in Figs. (\ref{Msqfig1}) and (\ref{Msqfig2}), the variations of the  MDMs obtained by using the $J_\mu^1$ and $J_\mu^3$ interpolating currents as functions of  the $M^2$ and $s_0$ parameters are  also presented. As can be seen from these figures, we have good stability of MDMs with respect to variations of the  $M^2$ and $s_0$ in their working regions. This is another requirement of the method that the physical quantities show weak dependence on the auxiliary parameters. 
%

\begin{table}[htp]
	\addtolength{\tabcolsep}{10pt}
	\caption{Working intervals of  $s_0$ and  $M^2$ as well as  the PC  and OPE convergence for the MDMs of $T_{QQ \bar q \bar s}$ and 
	$T_{QQ \bar s \bar s}$ states.}
	\label{parameter}
		\begin{center}
\begin{tabular}{l|ccccc}
                \hline\hline
                \\
~~State~~ & ~~Current~~ & $s_0$ (GeV$^2$)& 
$M^2$ (GeV$^2$) & ~~  PC ($\%$) ~~ & ~~  OPE Convergence 
 ($\%$) \\
\\
                                        \hline\hline
                                        \\
                        & $J^1_\mu$ & $20.5-22.5$ & $4.5-6.5$ & $33-59$ &  $2.2$  
                        \\
                        \\
                        & $J^2_\mu$ & $21.0-23.0$ & $4.5-6.5$ & $33-58$ &  $2.3$  \\
                        \\
~~$cc \bar{q} \bar{s}$~~  & $J^3_\mu$ & $20.5-22.5$ & $4.5-6.5$ & $34-60$ & $ 1.9$   \\
\\
                        & $J^4_\mu$ & $20.5-22.5$ & $4.5-6.5$ & $32-57$ &  $2.5$   \\
                        \\
                                       \hline\hline
                                       \\
                        & $J^3_\mu$ & $21.0-23.0$  & $4.5-6.5$ & $34-58$ & $2.0$  \\
                        \\
 ~~$cc \bar{s} \bar{s}$~~ & $J^4_\mu$ & $21.0-23.0$  & $4.5-6.5$ & $33-58$ &  $2.0$  \\
 \\
                                        \hline\hline
                                        \\
                        & $J^1_\mu$ & $115.0-119.0$ & $11.0-15.0$ & $35-62$ &  $1.8$  \\
                        \\
                        & $J^2_\mu$ & $115.0-119.0$ & $11.0-15.0$ & $34-61$ &  $1.8$   \\
                        \\
~~$bb \bar{q} \bar{s}$~~  & $J^3_\mu$ & $115.0-119.0$ &$11.0-15.0$ & $34-61$ & $1.7$   \\
\\
                        & $J^4_\mu$ & $115.0-119.0$ & $11.0-15.0$ & $34-59$ &  $1.7$   \\
                        \\
                                       \hline\hline
                                       \\
                        & $J^3_\mu$ & $115.5-119.5$ & $11.0-15.0$ & $33-60$ &  $1.8$  \\
                        \\
 ~~$bb \bar{s} \bar{s}$~~ & $J^4_\mu$ & $115.5-119.5$ & $11.0-15.0$ & $34-62$ & $ 1.5 $ \\
 \\
                                    \hline\hline
 \end{tabular}
\end{center}
\end{table}


Using all the input parameters and working windows for $M^2$ and $s_0$ we present the  final results for the MDMs of the  $T_{QQ \bar q \bar s}$ and $T_{QQ \bar s \bar s}$ states  in Table \ref{MDMres1}. The presented uncertainties  in the values are  initiated from the errors in the values of all  the input parameters as well as those errors coming from the   computations of the working intervals for the auxiliary parameters $M^2$ and $s_0$. Since the  spectroscopic parameters for $T_{QQ \bar s \bar s}$ obtained by using the  $J_\mu^1$ and $J_\mu^2$ currents are not available, the MDMs of these states could not be calculated.
As can be seen from the  Table \ref{MDMres1}, different currents used to study tetraquarks having the same quark content yield different results for their MDMs at all. This implies that the results of MDMs strongly depend on the structure and nature of the diquark and antidiquark configurations  with the same quark contents. A glance at this table tells us that the results obtained by using  $J_\mu^1$ and $J_\mu^2$ are close to/roughly consistent with each other within the presented errors for all the considered states.  This is the case also for the results obtained via $J_\mu^3$ and $J_\mu^4$.  But,  there is a considerable difference between these two sets: The currents  $J_\mu^3$ and $J_\mu^4$ produce large MDMs compared to the currents  $J_\mu^1$ and $J_\mu^2$.  As we mentioned previously, the currents $J_\mu^1$,  $J_\mu^2$,   $J_\mu^3$ and  $J_\mu^4$ contain the heavy axial vector  diquak-light scalar antidiquark, heavy scalar diquak-light axial vector  antidiquark, heavy tensor  diquak-light axial vector  antidiquark and heavy vector diquak-light pseudotensor antidiquark, respectively. It is evident that the MDMs obtained for each channel are large in the tensor/pseudotensor-axial vector/vector configurations compared to the scalar/axial vector-axial vector/scalar ones.
As  we also previously mentioned, all of these currents have the same quantum numbers and quark contents, hence,  in principle they should interpolate the same physical state at each channel.  But, as we saw,  the results are sensitive to the configurations and nature of the diquarks and antidiquarks that form the states under study.  
Although the PC is roughly the same in Table \ref{parameter} for all the currents at each channel in the working intervals of the $ s_0 $ and $ M^2 $,  it is small compared to the ordinary hadrons. Hence, a part of the differences among the results obtained using different currents can be related to this fact.
Therefore, as a result, the difference among the obtained results using  different currents can be attributed to different diquark-antidiquark structures of the currents  as well as systematic uncertainties of the method used.

\begin{table}[htp]
	\addtolength{\tabcolsep}{10pt}
	\caption{Numerical values of the MDMs for the $T_{QQ \bar q \bar s}$ and $T_{QQ \bar s \bar s}$ states, in units of nuclear magneton $\mu_N$.}
	\label{MDMres1}
		\begin{center}
\begin{tabular}{l|ccccccccccc}
	   \hline\hline\\
    Current&$T_{cc \bar u \bar s}$& $T_{cc \bar d \bar s}$& $T_{cc \bar s \bar s}$&$T_{bb \bar u \bar s}$& $T_{bb \bar d \bar s}$& $T_{bb \bar s \bar s}$\\
    \\
\hline\hline
\\
\\
 $J_\mu^1$     &$ 0.22 \pm 0.07$  &    $ 0.15 \pm 0.05$       & -  
 &$ -0.46 \pm 0.07$  &    $ -0.59 \pm 0.09$ &-     
 \\
 \\
\hline \hline
\\ 
\\
 $J_\mu^2$    &$ ~0.18 \pm 0.06$  &  $~0.36 \pm 0.11$         &  -
 &$ -0.34 \pm 0.12$  &    $ -0.68 \pm 0.24$ &  -     
 \\
 \\
\hline \hline
\\
\\
$J_\mu^3$     &$~1.09 \pm 0.36$  &   $~1.70 \pm 0.57$        &  $~1.52 \pm 0.46$ 
&$ -2.37 \pm 0.50$  &    $ -1.37 \pm 0.17$ &  $ -2.90 \pm 0.23$   \\
\\
\hline \hline
\\
\\
$J_\mu^4$    &$ ~1.04 \pm 0.35$ &    $~2.08 \pm 0.64$         &  $~1.90 \pm 0.56$ &
$ -2.90 \pm 0.36$  &    $ -1.45 \pm 0.33$ &  $ -3.20 \pm 0.48$   \\
\\
	   \hline\hline
\end{tabular}
\end{center}
\end{table}
%
%

As a final remark, we would like to briefly  discuss  how it is possible to  measure the MDMs of $T_{QQ \bar q \bar s}$ and $T_{QQ \bar s \bar s}$ states in future experiments. Although  the short lifetimes of doubly heavy tetraquark states make it difficult  to experimentally measure their MDMs directly, further accumulation of data in future may make this possible. $\Delta^+(1232)$ baryon has also a very short lifetime,  but its MDM is accessible from the $\gamma N \rightarrow  \Delta \rightarrow  \Delta \gamma \rightarrow \pi N \gamma $ processes \cite{Pascalutsa:2004je, Pascalutsa:2005vq, Pascalutsa:2007wb}. One approach for definition of the electromagnetic multipole moments of hadrons  is based on the soft photon emission off the hadrons (see for instance Ref.~\cite{Zakharov:1968fb}). The photon carries useful knowledge on the electromagnetic properties of the mother hadron. The matrix element for  such  radiative process can be parameterized in terms of the photon's energy, $E_\gamma$, 
\begin{align}
 M \sim A(E_\gamma)^{-1} + B(E_\gamma )^0 +\cdots,
\end{align}
where the charge, MDM and higher multipole moments contributions to the amplitude are denoted by $(E_\gamma )^{-1}$, $(E_\gamma)^0$ and $\cdots$, respectively.  By measuring the  decay width or cross-section of the radiative process of the doubly heavy tetraquarks  and neglecting from the small contributions of the terms linear or higher order in $E_\gamma$, one can extract the MDM of the corresponding state.

\section{Summary and Concluding Remarks}

 The mass and width of a particle are two main quantities, calculations of which and their comparisons  with the experimental data can provide us with  useful knowledge on the nature, inner structure and quantum numbers of the composite particles. In the case of the newly discovered multiquark exotic states  by different experiments, however, such comparisons have not  lead to  satisfactory results: Most of the discovered tetra and pentaquarks are not well-established and their internal organizations are not clear yet. For instance, for  the tetraquarks, both of  the molecular and compact tetraquark pictures well produce the experimental data on the related mass and width. Hence, investigation of different decays/interactions of these states to/with other known particles can play an important role. Determination of the electromagnetic properties of these hadrons can be very useful in this regard. 
 Motivated by this statement as well as the recent observation of the very narrow doubly-charmed $T^+_{cc}$ state by the LHCb collaboration,    in this article, we have explored the magnetic dipole moments of the doubly-heavy single/double strange $T_{QQ \bar q \bar s}$ and $T_{QQ \bar s \bar s}$ states  having  the spin-parity $ J^{P} = 1^{+}$ in the framework of the QCD light-cone sum rule method. 
  The magnetic dipole moment is the leading-order response of a hadronic  system to a soft external magnetic field. Therefore, it provides an excellent platform for investigation of the nature and   inner structure of hadrons as governed by the quark-gluon dynamics of QCD as well as their geometric shape.
  To calculate the magnetic dipole moments, different interpolating currents that can couple to the $T_{QQ \bar q \bar s}$ and $T_{QQ \bar s \bar s}$ states are considered. 
As is seen from Table \ref{MDMres1}, different interpolating currents lead to different predictions for the magnetic dipole moments of the $T_{QQ \bar q \bar s}$  and  $T_{QQ \bar s \bar s}$ states. 
The difference among the obtained results using the different currents can be attributed to different diquark-antidiquark structures of the currents as well as systematic uncertainties of the method used.

The results obtained in this study on the magnetic dipole moments of the doubly-heavy exotic tetraquark states in both the charmed-strange and bottom-strange channels  can be used in future experimental examinations of the multiquark states. Our results may be checked via other phenomenological approaches as well.

   \section*{ACKNOWLEDGEMENTS}
K. Azizi is grateful to Iran Science Elites Federation (Saramadan)
for the partial  financial support provided under the grant number ISEF/M/401385.

 \newpage
 
\appendix

\subsection*{Appendix  A: Feynman Diagrams}
In this appendix,we present some Feynman diagrams, which have been  taken into account in the calculation of the QCD side of the correlation function.
\begin{figure}[htp]\label{Feyndiag}
\subfloat[]{ \includegraphics[width=0.55\textwidth]{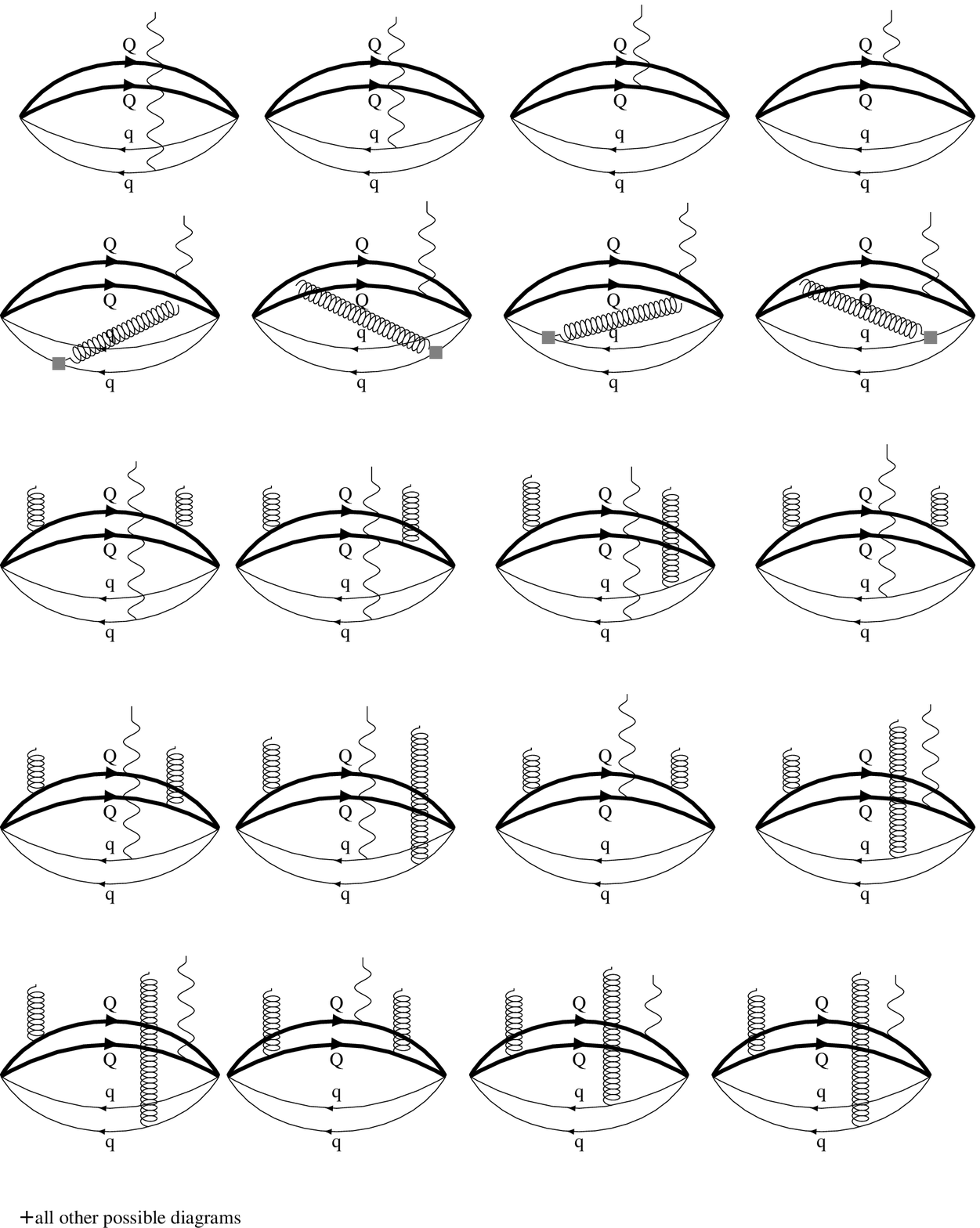}}\\
\subfloat[]{ \includegraphics[width=0.55\textwidth]{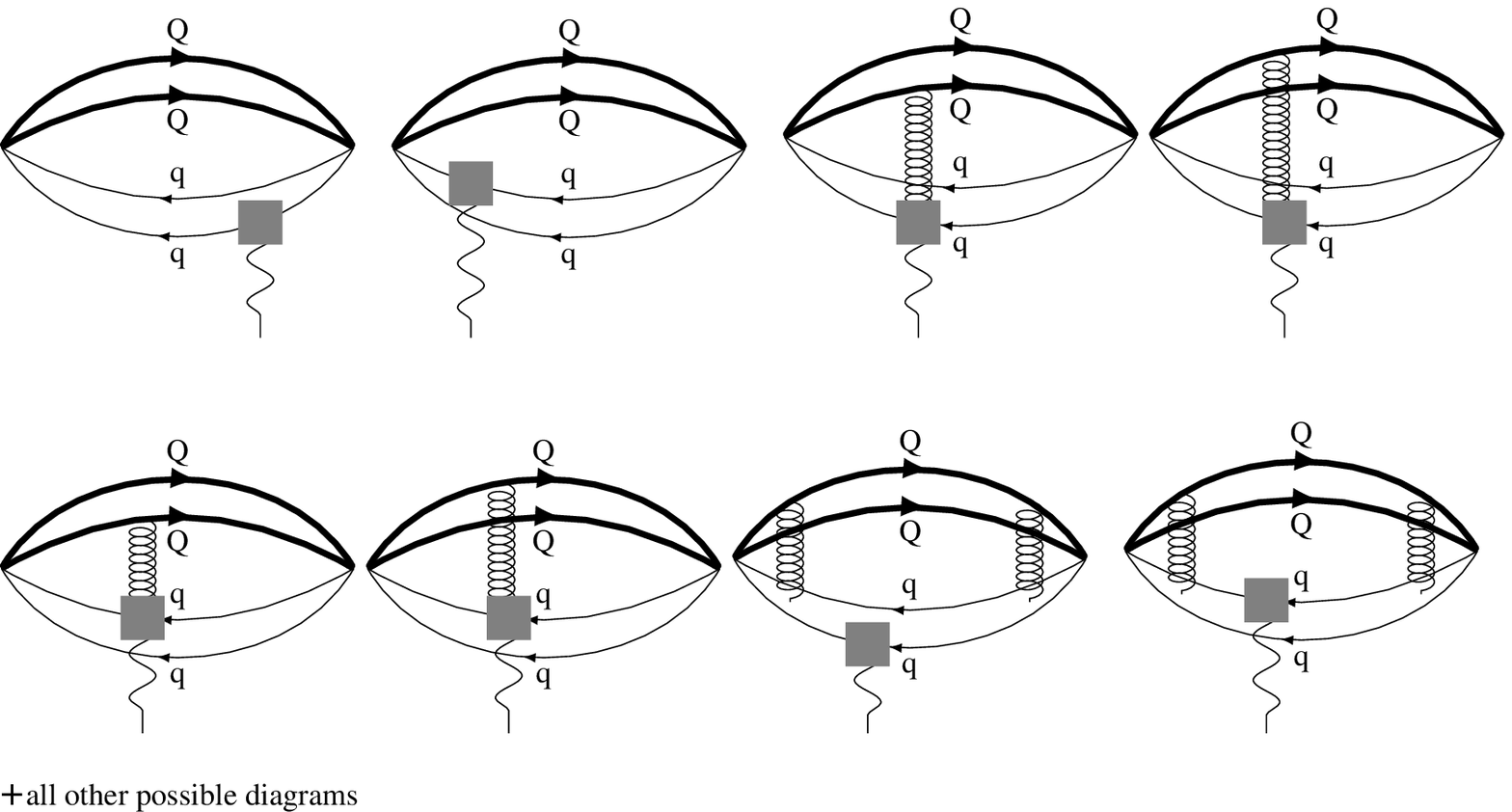}}
 \caption{ Feynman diagrams for the  magnetic dipole moments of the doubly-heavy tetraquark states. The heavy quark, light quark, photon and gluon lines are represented by  the thick, thin, wavy and curly lines, respectively. 
 Diagrams (a) correspond to the perturbative photon emission and  diagrams (b) represent the contributions coming from the distribution amplitudes of the photon.}
  \end{figure}

 \subsection*{Appendix  B: Explicit expression for \texorpdfstring{$\Delta_1 (M^2,s_0)$}{}}
 In this appendix, we present the explicit expression of the  $\Delta_1 (M^2,s_0)$ function for the MDMs of doubly-heavy tetraquark states:
\begin{align}
  \Delta_1 (M^2,s_0) &= \frac{m_{T_{QQ}}^2}{\lambda^2_{T_{QQ}}}\, e^{\frac{m_{T_{QQ}}^2}{M^2}}\Bigg\{ 
  -\frac {e_Q} {1310720 \pi^5} \Big[
   I[0, 5, 2, 2] - 3 I[0, 5, 2, 3] + 3 I[0, 5, 2, 4] - 
    I[0, 5, 2, 5] - 3 I[0, 5, 3, 2] \nonumber\\
  &+ 6 I[0, 5, 3, 3] - 
    3 I[0, 5, 3, 4] + 3 I[0, 5, 4, 2] - 3 I[0, 5, 4, 3] - 
    I[0, 5, 5, 2]\Big]
  \nonumber\\\
  &
+\frac {\langle \bar q q \rangle \langle \bar ss \rangle m_0^2} {  5308416  \pi^3}\Big[e_Q \Big(
   I[0, 1, 1, 0] +3 I[0, 1, 1, 1] -2 I[0, 1, 1, 2] + 
    2 I[0, 1, 2, 0] -3 I[0, 1, 2, 1] 
    + 
    I[0, 1, 3, 0]  \Big)\Big]\nonumber\\
  &
  +\frac{\langle g_s^2 G^2 \rangle \langle \bar ss \rangle } {4718592 \pi^3 }\Big[
   (e_q-12 e_Q)m_s\Big(I[0, 1, 2, 0] - 3 I[0, 1, 2, 1] + 3 I[0, 1, 2, 2] - 
       I[0, 1, 2, 3] - 3 I[0, 1, 3, 0]\Big)\Big]\nonumber\\
         &
        +\frac {\langle g_s^2 G^2 \rangle f_ {3 \gamma}} {221184 \pi^3}\Big[(e_q + e_s) m_Q^2 \Big(
   I[0, 1, 1, 0] - 2 I[0, 1, 1, 1] + I[0, 1, 1, 2] - 
    2 I[0, 1, 2, 0] + 2 I[0, 1, 2, 1] \nonumber\\
    &+ 
    I[0, 1, 3, 0]  \Big)\Big]\varphi_a (u_ 0)\nonumber\\
       &
       +\frac {m_0^2 \Big(e_s \langle \bar q q \rangle -e_u \langle \bar ss \rangle\Big) m_s} {2654208 \pi^3}\Big[
   I[0, 2, 1, 0] +3 I[0, 2, 1, 1] -2 I[0, 2, 1, 2] + 
    2 I[0, 2, 2, 0] -3 I[0, 2, 2, 1] \nonumber\\
    &
    + 
    I[0, 1, 3, 0] \Big]
  \nonumber\\
      &
  -\frac{\langle g_s^2 G^2 \rangle} {1179648 \pi^5 }\Big[
   3 e_Q \Big(I[0, 3, 2, 0] - 3 I[0, 3, 2, 1] + 3 I[0, 3, 2, 2] - 
       I[0, 3, 2, 3] - 3 I[0, 3, 3, 0] \nonumber\\
       &+ 6 I[0, 3, 3, 1] - 
       3 I[0, 3, 3, 2] + 3 I[0, 3, 4, 0] - 3 I[0, 3, 4, 1] - 
       I[0, 3, 5, 0]\Big) + 
    2 e_s \Big(4  m_Q^2 \big (I[0, 2, 1, 1] \nonumber\\
          &- 2 I[0, 2, 1, 2] + 
          I[0, 2, 1, 3] - 2 I[0, 2, 2, 1] + 2 I[0, 2, 2, 2] + 
          I[0, 2, 3, 1]\big) + I[0, 3, 2, 0] - 3 I[0, 3, 2, 1]\nonumber\\
       & + 
       3 I[0, 3, 2, 2] - I[0, 3, 2, 3] - 3 I[0, 3, 3, 0] + 
       6 I[0, 3, 3, 1] - 3 I[0, 3, 3, 2] + 3 I[0, 3, 4, 0] - 
       3 I[0, 3, 4, 1] \nonumber\\
       & - I[0, 3, 5, 0]\Big) + 
    e_q \Big(-4  m_Q^2 \big(I[0, 2, 1, 1] - 2 I[0, 2, 1, 2] + 
           I[0, 2, 1, 3] - 2 I[0, 2, 2, 1] + 2 I[0, 2, 2, 2]\nonumber\\
       & + 
           I[0, 2, 3, 1]\big) - I[0, 3, 2, 0] + 3 I[0, 3, 2, 1] - 
        3 I[0, 3, 2, 2] + I[0, 3, 2, 3] + 3 I[0, 3, 3, 0] - 
        6 I[0, 3, 3, 1] \nonumber\\
       & + 3 I[0, 3, 3, 2] - 3 I[0, 3, 4, 0] + 
        3 I[0, 3, 4, 1] + I[0, 3, 5, 0]\Big)\Big]
        \nonumber\\
        &
        +\frac { \Big(\langle \bar q q \rangle-2 \langle \bar ss \rangle\Big)} {8192 \pi^3} \Big[e_Q m_s\Big(
   I[0, 3, 2, 0] - 3 I[0, 3, 2, 1] + 3 I[0, 3, 2, 2] - 
    I[0, 3, 2, 3] - 3 I[0, 3, 3, 0] \nonumber\\
        & + 6 I[0, 3, 3, 1]- 
    3 I[0, 3, 3, 2] + 3 I[0, 3, 4, 0] - 3 I[0, 3, 4, 1] - 
    I[0, 3, 5, 0]\Big)\Big]\nonumber\\
      &
    -\frac { \langle \bar qq \rangle} {16384 \pi^3} e_q\, m_s\, I_3[\mathcal S]\, I[0, 3, 5, 0]+\frac {f_ {3 \gamma}} {131072 \pi^3}\Big (e_s I_ 1[\mathcal V] + 
   e_q I_ 2[\mathcal V]\Big) I[0, 4, 5, 0]
    \Bigg\},
\end{align}
where $e_{q}$, $m_{q}$,  and $\langle \bar q q \rangle $ are  the electric charge, mass, and condensates of the corresponding light-quark, respectively.  Here $u_0= \frac{M_1^2}{M_1^2+M_2^2} $, and  $ \frac{1}{M^2}= \frac{M_1^2+M_2^2}{M_1^2 M_2^2}$ with $ M_1^2 $ and $ M_2^2 $ being  the Borel parameters in the initial and final states, respectively.   As the initial and final states are the same particles, we set $ M_1^2= M_2^2= 2 M^2$. 
  
The functions~$I[n,m,l,k]$, $I_1[\mathcal{F}]$,~$I_2[\mathcal{F}]$, and ~$I_3[\mathcal{F}]$ appeared in the expression of  $\Delta_1 (M^2,s_0)$ are
given  as:
\begin{align}
 I[n,m,l,k]&= \int_{4m_b^2}^{s_0} ds \int_{0}^1 dt \int_{0}^1 dw~ e^{-s/M^2}~
 s^n\,(s-4m_b^2)^m\,t^l\,w^k,\nonumber\\
 I_1[\mathcal{F}]&=\int D_{\alpha_i} \int_0^1 dv~ \mathcal{F}(\alpha_{\bar q},\alpha_q,\alpha_g)
 \delta'(\alpha_ q +\bar v \alpha_g-u_0),\nonumber\\
  I_2[\mathcal{F}]&=\int D_{\alpha_i} \int_0^1 dv~ \mathcal{F}(\alpha_{\bar q},\alpha_q,\alpha_g)
 \delta'(\alpha_{\bar q}+ v \alpha_g-u_0),\nonumber\\
     I_3[\mathcal{F}]&=\int D_{\alpha_i} \int_0^1 dv~ \mathcal{F}(\alpha_{\bar q},\alpha_q,\alpha_g)
 \delta(\alpha_{\bar q}+ v \alpha_g-u_0),\nonumber
 \end{align}
 where $\mathcal{F}$ denotes the corresponding photon wave function.

  \section*{Appendix C: Distribution Amplitudes of the on-shell photon }
In this appendix, we present the  the matrix elements $\langle \gamma(q)\vel \bar{q}(x) \Gamma_i q(0) \ver 0\rangle$  
and $\langle \gamma(q)\vel \bar{q}(x) \Gamma_i G_{\mu\nu}q(0) \ver 0\rangle$ entering  the calculations \cite{Ball:2002ps} :
\begin{eqnarray*}
\label{esbs14}
&&\langle \gamma(q) \vert  \bar q(x) \gamma_\mu q(0) \vert 0 \rangle
= e_q f_{3 \gamma} \left(\varepsilon_\mu - q_\mu \frac{\varepsilon
x}{q x} \right) \int_0^1 du e^{i \bar u q x} \psi^v(u)
\nonumber \\
&&\langle \gamma(q) \vert \bar q(x) \gamma_\mu \gamma_5 q(0) \vert 0
\rangle  = - \frac{1}{4} e_q f_{3 \gamma} \epsilon_{\mu \nu \alpha
\beta } \varepsilon^\nu q^\alpha x^\beta \int_0^1 du e^{i \bar u q
x} \psi^a(u)
\nonumber \\
&&\langle \gamma(q) \vert  \bar q(x) \sigma_{\mu \nu} q(0) \vert  0
\rangle  = -i e_q \langle \bar q q \rangle (\varepsilon_\mu q_\nu - \varepsilon_\nu
q_\mu) \int_0^1 du e^{i \bar u qx} \left(\chi \varphi_\gamma(u) +
\frac{x^2}{16} \mathbb{A}  (u) \right) \nonumber \\ 
&&-\frac{i}{2(qx)}  e_q \bar qq \left[x_\nu \left(\varepsilon_\mu - q_\mu
\frac{\varepsilon x}{qx}\right) - x_\mu \left(\varepsilon_\nu -
q_\nu \frac{\varepsilon x}{q x}\right) \right] \int_0^1 du e^{i \bar
u q x} h_\gamma(u)
\nonumber \\
&&\langle \gamma(q) | \bar q(x) g_s G_{\mu \nu} (v x) q(0) \vert 0
\rangle = -i e_q \langle \bar q q \rangle \left(\varepsilon_\mu q_\nu - \varepsilon_\nu
q_\mu \right) \int {\cal D}\alpha_i e^{i (\alpha_{\bar q} + v
\alpha_g) q x} {\cal S}(\alpha_i)
\nonumber \\
&&\langle \gamma(q) | \bar q(x) g_s \tilde G_{\mu \nu}(v
x) i \gamma_5  q(0) \vert 0 \rangle = -i e_q \langle \bar q q \rangle \left(\varepsilon_\mu q_\nu -
\varepsilon_\nu q_\mu \right) \int {\cal D}\alpha_i e^{i
(\alpha_{\bar q} + v \alpha_g) q x} \tilde {\cal S}(\alpha_i)
\nonumber \\
&&\langle \gamma(q) \vert \bar q(x) g_s \tilde G_{\mu \nu}(v x)
\gamma_\alpha \gamma_5 q(0) \vert 0 \rangle = e_q f_{3 \gamma}
q_\alpha (\varepsilon_\mu q_\nu - \varepsilon_\nu q_\mu) \int {\cal
D}\alpha_i e^{i (\alpha_{\bar q} + v \alpha_g) q x} {\cal
A}(\alpha_i)
\nonumber \\
&&\langle \gamma(q) \vert \bar q(x) g_s G_{\mu \nu}(v x) i
\gamma_\alpha q(0) \vert 0 \rangle = e_q f_{3 \gamma} q_\alpha
(\varepsilon_\mu q_\nu - \varepsilon_\nu q_\mu) \int {\cal
D}\alpha_i e^{i (\alpha_{\bar q} + v \alpha_g) q x} {\cal
V}(\alpha_i) \nonumber\\
&& \langle \gamma(q) \vert \bar q(x)
\sigma_{\alpha \beta} g_s G_{\mu \nu}(v x) q(0) \vert 0 \rangle  =
e_q \langle \bar q q \rangle \left\{
        \left[\left(\varepsilon_\mu - q_\mu \frac{\varepsilon x}{q x}\right)\left(g_{\alpha \nu} -
        \frac{1}{qx} (q_\alpha x_\nu + q_\nu x_\alpha)\right) \right. \right. q_\beta
\nonumber \\
 && -
         \left(\varepsilon_\mu - q_\mu \frac{\varepsilon x}{q x}\right)\left(g_{\beta \nu} -
        \frac{1}{qx} (q_\beta x_\nu + q_\nu x_\beta)\right) q_\alpha
-
         \left(\varepsilon_\nu - q_\nu \frac{\varepsilon x}{q x}\right)\left(g_{\alpha \mu} -
        \frac{1}{qx} (q_\alpha x_\mu + q_\mu x_\alpha)\right) q_\beta
\nonumber \\
 &&+
         \left. \left(\varepsilon_\nu - q_\nu \frac{\varepsilon x}{q.x}\right)\left( g_{\beta \mu} -
        \frac{1}{qx} (q_\beta x_\mu + q_\mu x_\beta)\right) q_\alpha \right]
   \int {\cal D}\alpha_i e^{i (\alpha_{\bar q} + v \alpha_g) qx} {\cal T}_1(\alpha_i)
\nonumber \\
 &&+
        \left[\left(\varepsilon_\alpha - q_\alpha \frac{\varepsilon x}{qx}\right)
        \left(g_{\mu \beta} - \frac{1}{qx}(q_\mu x_\beta + q_\beta x_\mu)\right) \right. q_\nu
\nonumber \\ &&-
         \left(\varepsilon_\alpha - q_\alpha \frac{\varepsilon x}{qx}\right)
        \left(g_{\nu \beta} - \frac{1}{qx}(q_\nu x_\beta + q_\beta x_\nu)\right)  q_\mu
\nonumber \\ && -
         \left(\varepsilon_\beta - q_\beta \frac{\varepsilon x}{qx}\right)
        \left(g_{\mu \alpha} - \frac{1}{qx}(q_\mu x_\alpha + q_\alpha x_\mu)\right) q_\nu
\nonumber \\ &&+
         \left. \left(\varepsilon_\beta - q_\beta \frac{\varepsilon x}{qx}\right)
        \left(g_{\nu \alpha} - \frac{1}{qx}(q_\nu x_\alpha + q_\alpha x_\nu) \right) q_\mu
        \right]      
    \int {\cal D} \alpha_i e^{i (\alpha_{\bar q} + v \alpha_g) qx} {\cal T}_2(\alpha_i)
\nonumber \\
&&+\frac{1}{qx} (q_\mu x_\nu - q_\nu x_\mu)
        (\varepsilon_\alpha q_\beta - \varepsilon_\beta q_\alpha)
    \int {\cal D} \alpha_i e^{i (\alpha_{\bar q} + v \alpha_g) qx} {\cal T}_3(\alpha_i)
\nonumber \\ &&+
        \left. \frac{1}{qx} (q_\alpha x_\beta - q_\beta x_\alpha)
        (\varepsilon_\mu q_\nu - \varepsilon_\nu q_\mu)
    \int {\cal D} \alpha_i e^{i (\alpha_{\bar q} + v \alpha_g) qx} {\cal T}_4(\alpha_i)
                        \right\},
\end{eqnarray*}
where the measure ${\cal D} \alpha_i$ is defined as
\begin{eqnarray*}
\label{nolabel05}
\int {\cal D} \alpha_i = \int_0^1 d \alpha_{\bar q} \int_0^1 d
\alpha_q \int_0^1 d \alpha_g \delta(1-\alpha_{\bar
q}-\alpha_q-\alpha_g)~.\nonumber
\end{eqnarray*}
Here, $\varphi_\gamma(u)$ is the DA of leading twist-2, $\psi^v(u)$,
$\psi^a(u)$, ${\cal A}(\alpha_i)$ and ${\cal V}(\alpha_i)$, are the twist-3 amplitudes, and
$h_\gamma(u)$, $\mathbb{A}(u)$, ${\cal S}(\alpha_i)$, ${\cal{\tilde S}}(\alpha_i)$, ${\cal T}_1(\alpha_i)$, ${\cal T}_2(\alpha_i)$, ${\cal T}_3(\alpha_i)$ 
and ${\cal T}_4(\alpha_i)$ are the
twist-4 photon DAs.

The explicit expressions of the DAs that are entered into the matrix elements above are  given as follows:
\begin{eqnarray}
\varphi_\gamma(u) &=& 6 u \bar u \left( 1 + \varphi_2(\mu)
C_2^{\frac{3}{2}}(u - \bar u) \right),
\nonumber \\
\psi^v(u) &=& 3 \left(3 (2 u - 1)^2 -1 \right)+\frac{3}{64} \left(15
w^V_\gamma - 5 w^A_\gamma\right)
                        \left(3 - 30 (2 u - 1)^2 + 35 (2 u -1)^4
                        \right),
\nonumber \\
\psi^a(u) &=& \left(1- (2 u -1)^2\right)\left(5 (2 u -1)^2 -1\right)
\frac{5}{2}
    \left(1 + \frac{9}{16} w^V_\gamma - \frac{3}{16} w^A_\gamma
    \right),
\nonumber \\
h_\gamma(u) &=& - 10 \left(1 + 2 \kappa^+\right) C_2^{\frac{1}{2}}(u
- \bar u),
\nonumber \\
\mathbb{A}(u) &=& 40 u^2 \bar u^2 \left(3 \kappa - \kappa^+
+1\right)  +
        8 (\zeta_2^+ - 3 \zeta_2) \left[u \bar u (2 + 13 u \bar u) \right.
\nonumber \\ && + \left.
                2 u^3 (10 -15 u + 6 u^2) \ln(u) + 2 \bar u^3 (10 - 15 \bar u + 6 \bar u^2)
        \ln(\bar u) \right],
\nonumber \\
{\cal A}(\alpha_i) &=& 360 \alpha_q \alpha_{\bar q} \alpha_g^2
        \left(1 + w^A_\gamma \frac{1}{2} (7 \alpha_g - 3)\right),
\nonumber \\
{\cal V}(\alpha_i) &=& 540 w^V_\gamma (\alpha_q - \alpha_{\bar q})
\alpha_q \alpha_{\bar q}
                \alpha_g^2,
\nonumber \\
{\cal T}_1(\alpha_i) &=& -120 (3 \zeta_2 + \zeta_2^+)(\alpha_{\bar
q} - \alpha_q)
        \alpha_{\bar q} \alpha_q \alpha_g,
\nonumber \\
{\cal T}_2(\alpha_i) &=& 30 \alpha_g^2 (\alpha_{\bar q} - \alpha_q)
    \left((\kappa - \kappa^+) + (\zeta_1 - \zeta_1^+)(1 - 2\alpha_g) +
    \zeta_2 (3 - 4 \alpha_g)\right),
\nonumber \\
{\cal T}_3(\alpha_i) &=& - 120 (3 \zeta_2 - \zeta_2^+)(\alpha_{\bar
q} -\alpha_q)
        \alpha_{\bar q} \alpha_q \alpha_g,
\nonumber \\
{\cal T}_4(\alpha_i) &=& 30 \alpha_g^2 (\alpha_{\bar q} - \alpha_q)
    \left((\kappa + \kappa^+) + (\zeta_1 + \zeta_1^+)(1 - 2\alpha_g) +
    \zeta_2 (3 - 4 \alpha_g)\right),\nonumber \\
{\cal S}(\alpha_i) &=& 30\alpha_g^2\{(\kappa +
\kappa^+)(1-\alpha_g)+(\zeta_1 + \zeta_1^+)(1 - \alpha_g)(1 -
2\alpha_g)\nonumber +\zeta_2[3 (\alpha_{\bar q} - \alpha_q)^2-\alpha_g(1 - \alpha_g)]\},\nonumber \\
\tilde {\cal S}(\alpha_i) &=&-30\alpha_g^2\{(\kappa -\kappa^+)(1-\alpha_g)+(\zeta_1 - \zeta_1^+)(1 - \alpha_g)(1 -
2\alpha_g)\nonumber +\zeta_2 [3 (\alpha_{\bar q} -\alpha_q)^2-\alpha_g(1 - \alpha_g)]\}.
\end{eqnarray}
The numerical values of the parameters used in the DAs are: $\varphi_2(1~GeV) = 0$, 
$w^V_\gamma = 3.8 \pm 1.8$, $w^A_\gamma = -2.1 \pm 1.0$, $\kappa = 0.2$, $\kappa^+ = 0$, $\zeta_1 = 0.4$, and $\zeta_2 = 0.3$.

\bibliography{article.bib}

\end{document}